\documentclass[pre,aps,twocolumn,showpacs,floatfix,10pt,superscriptaddress]{revtex4-2}
\usepackage{amsmath}
\usepackage{amsfonts}
\usepackage{amssymb}
\usepackage{graphicx}
\usepackage{bm}
\usepackage{color}
\usepackage{ulem}
 
 \usepackage{epsfig}
\usepackage{graphicx}
\usepackage{chngcntr}
\counterwithout{figure}{section}
\usepackage{rotating}
\usepackage{amssymb}
\usepackage{amsmath}
\usepackage{physics}
\usepackage{multirow}
\usepackage{float}
\usepackage{caption, subcaption}

\begin{document}

\title{Phase transitions in XY models with randomly oriented crystal fields}
\author{Sumedha,}
\email{sumedha@niser.ac.in}
\affiliation{School of Physical Sciences, National Institute of Science Education and Research,Bhubaneswar,P.O. Jatni,Khurda,Odisha,India 752050}
\affiliation{Homi Bhabha National Institute, Training School Complex, Anushakti Nagar, Mumbai,India 400094}
\author{Mustansir Barma}
\email{barma@tifrh.res.in}
\affiliation{TIFR Centre for Interdisciplinary Sciences, Tata Institute of Fundamental Research, Gopanpally, Hyderabad, India  500046}

\begin{abstract}
We obtain a representation of the free energy of an XY model on a fully connected graph with spins subjected to a random crystal field of strength $D$ and with random orientation $\alpha$. Results are obtained for an arbitrary probability distribution of the disorder using large deviation theory, for any $D$. We show that the critical temperature is insensitive to the nature and strength of the distribution $p(\alpha)$, for a large family of distributions which includes  quadriperiodic distributions, with $p(\alpha)=p(\alpha+\frac{\pi}{2})$, which includes the uniform and symmetric bimodal distributions. The specific heat vanishes as temperature $T \rightarrow 0$ if $D$ is infinite, but approaches a constant if $D$ is finite. We also studied the effect of asymmetry on a bimodal distribution of the orientation of the random crystal field and obtained the phase diagram comprising four phases: a mixed phase (in which spins are canted at angles which depend on the degree of disorder), an $x$-Ising phase, a $y$-Ising  phase and a paramagnetic phase, all of which meet at a tetra-critical point. The canted mixed phase is present for all finite $D$, but vanishes when $D \rightarrow \infty$.

\end{abstract}

\date{\today}
\maketitle

\section{Introduction}

Randomly anisotropic crystal fields play an important role in determining the magnetic properties of amorphous magnetic materials. In the random anisotropy model (RAM) \cite{harris},  each spin is subjected to a local anisotropy with random orientation in addition to the usual spin exchange interaction. While longitudinal random anisotropy has no effect on Ising spins, for vector spins it competes with the ferromagnetic exchange energy in determining the state of the system. The model provides a theoretical basis for understanding magnetic properties of many  amorphous binary alloys \cite{krey,alben, herzer, dudka}, nanocrystalline \cite{herzer,nano,alvarez} and molecular \cite{molecular} magnets. 

The RAM can be defined for vector spins of any dimensionality $m \ge 2$, but in this paper we study only XY spins, corresponding to $m=2$. In the limit of infinite strength of a crystal field oriented randomly, the model reduces to a quenched random-bond Ising model with correlated random couplings \cite{lubensky}, raising the possibility of a spin glass phase in the RAM. In this limit, the model was conjectured to belong to the same universality class as the Edwards-Anderson Ising spin glass model \cite{bray}.

Since its introduction, the model has been studied  using many techniques such as mean field theory \cite{callen,patterson,dv}, variational methods \cite{variational}, field theories \cite{mukamel,dudka,pelcovits,shapoval,rg}, and Monte Carlo simulations \cite{simulation1, simulation2,simulation3,marinari}. The infinite crystal field limit has been studied extensively \cite{lubensky,dv,jayaprakash, fischer,marinari,fisch} both analytically and through simulations using the mapping to random-bond Ising models \cite{lubensky}. Most $\epsilon$-expansion and Monte Carlo studies in three dimensions have been inconclusive in determining the nature of the low temperature phase. An intriguing feature of all the $\epsilon$- expansion based renormalisation group studies is that the stable fixed points cannot be reached from the initial conditions given by unrenormalized physically relevant effective Hamiltonians \cite{shapoval,referee}. In general, the distribution of the random axis plays a crucial role in determining the low energy configurations and phase transitions. 

We study the effect of random crystal field anisotropy on XY spins (RCXY) on a fully connected graph, for any distribution of the orientation of the crystal field axis and any strength of the crystal field $D$, using large deviation theory (LDT) \cite{dembo,touchette}. In recent related work on fully connected graphs, LDT was used to perform the disorder averaging for discrete-spin random-field problems \cite{lowe,sumedhasingh, sumedhajana,kistler}. For vector spins, LDT was used to solve the problem in the pure case \cite{kirkpatrick}, and more recently to study XY models in random magnetic fields \cite{sumedhabarma}. 

In this paper we use LDT to obtain the phase diagram and low temperature properties of the XY model with quenched uniform and bimodal distributions for the orientation of the crystal field. Earlier the model had been solved in the case $D = \infty$ \cite{dv}. Our solution for arbitrary $D$ brings out an unexpected invariance of the critical temperature $T_c$: For a large family of  distribution functions of the orientation (which includes the uniform and symmetric bimodal cases) there is a continuous transition at temperature $T_c = 1/2$, which coincides with $T_c$ for the pure XY model on a fully connected graph, even though the nature of the ordered phase depends on the details of the disorder distribution. Below we briefly discuss the two cases studied in this paper, namely the uniform and bimodal distributions of random orientations.

In amorphous alloys, the absence of the crystalline order implies there is no preferred direction for the crystal field and the system is often modelled as the RCXY model with a uniform distribution of the random axis orientation. In this case, we find that the $T = 0$ magnetization decreases as $D$ increases, approaching a finite value $2/\pi$ as $D \rightarrow \infty$ with a correction proportional to $1/D$ for large but finite $D$. Further the specific heat vanishes if $D \rightarrow \infty$, in agreement with earlier results \cite{dv}, but we show that it approaches a constant as $T \rightarrow 0$ for finite values of $D$.

We also study an asymmetric bimodal distribution of the orientation, with the crystal field pointing randomly along $x$ and $y$ directions on different fractions of sites, interpolating between the pure case and the quadriperiodic bimodal distribution. An interesting phase diagram ensues with three ordered phases : two phases where the magnetization is along only one of the $x$ or $y$ directions, and a mixed phase with a magnetization that is canted in two different directions. Four critical curves meet at a tetra-critical point which occurs for all asymmetric bimodal distributions of the random crystal field orientation. Tetra-critical points have also arisen in several other contexts where there are two order paramters, for instance anisotropic anti-ferromagnets \cite{kosterlitz}, alloys of materials with different anisotropies \cite{aharony} , strongly correlated SO(5) superconductors \cite{zhang,murakami} and other strongly correlated theories like quantum chromodynamics (QCD) \cite{sannino}. In the RCXY model under study here, the tetracritical point originates from the asymmetric discrete distribution of the crystal field, which produces an $x-y$ asymmetry between order parameters.

The plan of the paper is as follows : In Section \ref{model} we define the model and derive the expression of the rate function 
using the large-deviation theory, for any distribution of the quenched random orientation of the crystal field. We study the phase diagram and low temperature phase for the case of uniform distribution (Sec. \ref{flatd1}) and bimodal distribution (Sec. \ref{bimodald}), obtaining a closed form expression for the rate function as $D \rightarrow \infty$, and an expression for large $D$, in powers of $1/D$. We study finite $D$ via Taylor expansion of the rate function. In Section \ref{discuss} we discuss the main results of the paper and some future directions.


\section{Random crystal field XY model}
\label{model}

 The Hamiltonian of the model on a fully connected graph is
\begin{equation}
H=-\frac{J}{2 N} (\sum_{i=1}^N \vec{s_i})^2 -D\sum_{i=1}^N (\vec{n_i}. \vec{s_i})^2
\label{Hxy}
\end{equation}
where $s_i$ are $m$-component vector spins in general. For $m=2$ (XY model), they can be represented as $s_i= cos \theta_i \hat{i}+  \sin \theta_i \hat{j}$. Here $\theta_i$ is a random variable chosen uniformly from the interval $[0,2 \pi]$, $D$ is the crystal field strength, $J$ is the coupling which we take to be $1$ and ${\hat{n}}_i = \cos \alpha_i \hat{i}+\sin \alpha_i \hat{j}$ is the site dependent direction of the crystal field. The coupling between pairs of spins has been set equal to unity. The Hamiltonian depends only on the orientation of the crystal field and hence we need to consider $\alpha$ only on the half circle ($\alpha$ and $\pi+\alpha$ are equivalent). The direction of the crystal field at each site is chosen randomly 
and frozen ; each $\alpha_i$ is an i.i.d chosen from a specified  distribution, $p(\alpha)$. The ferromagnetic coupling term in Eq. \ref{Hxy} tries to align spins in the same direction while the crystal field term tries to align spins with their random anisotropy axis (see Fig. 1), leading to frustration. We take $D$ to be positive, except for the pure case (no disorder), we allow either sign.

We study different forms of $p(\alpha)$ and their consequences in the subsequent sections. We use large deviation theory to perform the quenched disorder average and obtain the free energy of the model defined by Eq. \ref{Hxy} as explained below.
\begin{figure}
         \includegraphics[width=0.45\textwidth]{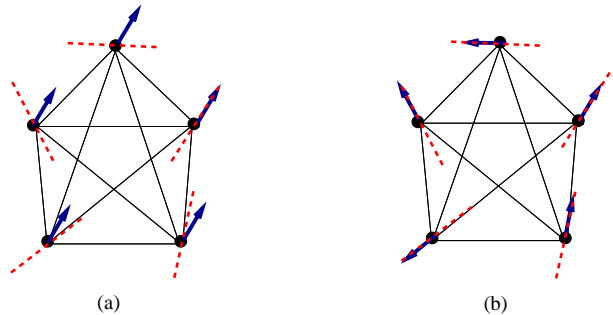}
             \caption{Low energy states with (a) $J \rightarrow \infty$ and (b) $D \rightarrow \infty$ on a fully connected graph with $N=5$ (five spins). Spins are represented by blue arrows and the random anisotropy axes by dotted red lines. In (a) the spins align with each other while in (b) the spins align with the random anisotropy axes.}
             \label{model}
             \end{figure}

\subsection*{Calculation of the free energy functional using LDP}
Consider any random configuration $C_N$ of $N$ spins with $x_1 =\sum_{i=1}^N \cos \theta_i/N$ and $x_2= \sum_{i=1}^N \sin \theta_i/N$. The probability of occurence of this configuration $P_{H,\beta}$ is proportional to $\exp(-\beta H)$, where $\beta=1/T$. The random variables $(\sum_{i=1}^N \cos \theta_i,\sum_{i=1}^N \sin \theta_i)$ satisfy the Large Deviation Principle (LDP) \cite{dembo,touchette,hollander} w.r.t to $P_{H,\beta}$. This implies that there exists a rate function $I(x_1,x_2)$ such that
\begin{equation}
P_{H,\beta}(C_N:x_1,x_2) \sim \exp(-N I(x_1,x_2))
\label{ldp}
\end{equation}
The rate function $I(x_1,x_2)$ is like the generalized free energy functional in that its minima give the free energy of the system. Recently the rate function was calculated exactly for discrete spin models with quenched random fields such as the random field Ising model and the random crystal field Blume-Capel model 
defined on a fully connected graph. It was shown that the rate-function when expanded in a power series is like a Landau free energy and hence can be used to extract the phase transitions in the system \cite{lowe,sumedhasingh,sumedhajana}. The extension of the method to vector spins, outlined below, was carried out for the random field XY model \cite{sumedhabarma}. 

There are two principal steps.
\begin{itemize}
\item Using the G{\"a}rtner-Ellis theorem \cite{dembo,touchette} and the law of large numbers, we first calculate the rate function $R(x_1,x_2)$ associated with the  the non-interacting part of the Hamiltonian  in Eq. \ref{Hxy} denoted by $H_{ni}$ and given by
\begin{equation}
H_{ni} = -D \sum_{i=1}^N (\cos \theta \cos \alpha_i+\sin \theta \sin \alpha_i)^2
\label{hni}
\end{equation}
Then $R(x_1,x_2)$ is defined through
\begin{equation}
P_{H_{ni},\beta}(C_N:x_1,x_2) \sim \exp(-N R(x_1,x_2))
\label{RHni}
\end{equation}
As we will see later in this section, the rate function $R(x_1,x_2)$ becomes independent of the specific realization of the disorder and depends only on $p(\alpha)$ in the limit $N \rightarrow \infty$.
\item The probability $P_{H,\beta}(C_N:x_1,x_2)$ for the Hamiltonian in Eq. \ref{Hxy} is proportional to $\int_A e^{N f(x_1,x_2)} P_{H_{ni},\beta}$, where $A$ is the subset of the all possible configurations, with a given $(x_1,x_2)$. The function $f(x_1,x_2)= \beta (x_1^2+x_2^2)/2$. The tilted large deviation principle \cite{hollander} then connects the two rate functions $I(x_1,x_2)$ and $R(x_1,x_2)$ through the relation
\begin{align}
I(x_1,x_2) &= R(x_1,x_2)-\frac{\beta x_1^2}{2} - \frac{\beta x_2^2}{2}\nonumber\\
&-\inf_{y_1,y_2} \left( R(y_1,y_2)-\frac{\beta y_1^2}{2} - \frac{\beta y_2^2}{2}\right)
\end{align}
The probability measure $P_{H,\beta}(C_N:x_1,x_2)$ is the tilted version of $P_{H_{ni},\beta}(C_N:x_1,x_2)$. 
\end{itemize}

Let us first calculate the rate function $R(x_1,x_2)$. Using the  G{\"a}rtner Ellis theorem it can be written as 
\begin{equation}
R(x_1,x_2) =  \sup_{y_1,y_2}  \{x_1 y_1 +x_2 y_2 -\Lambda(y_1,y_2)\}
\label{rfR}
\end{equation}
provided that the scaled cumulant generating function $\Lambda(y_1,y_2)= \lim_{N \rightarrow \infty} \Lambda_N(y_1,y_2)/N$, 
is differentiable \cite{dembo,touchette}. We calculate $\Lambda(y_1,y_2)$ 
for arbitrary distribution of the crystal field and show that it is differentiable. 

The function $\Lambda_N(y_1,y_2)$ is the log cumulant generating function for the probability distribution $P_{H_{ni},\beta}$
\begin{equation}
\Lambda_N(y_1,y_2) = \log \left\langle \exp( y_1 \sum_{i=1}^N \cos \theta_i + y_2 \sum_{i=1}^N \sin \theta_i) \right\rangle_Q
\end{equation}
Here $\langle ... \rangle_Q$ represents the expectation value w.r.t. 
the probability distribution $Q \propto e^{-\beta H_{ni}}$, which is a product measure over the probability distributions $Q_i$ for the non-interacting spins. Since $Q_i \propto \exp(\beta D \cos^2(\theta-\alpha_i))$, we obtain
\begin{align}
\Lambda(y_1,y_2) &= \lim_{N \rightarrow \infty} \frac{1}{N} \sum_{i=1}^N \log S_i
\label{lambdainf}
\end{align}
where 
\begin{equation}
S_i=\frac{1}{\tilde{N}} \int_0^{2 \pi} d \theta \exp(\beta D \cos^2 (\theta-\alpha_i) +y_1 \cos \theta +y_2 \sin \theta)
\label{S}
\end{equation}
Here $\tilde{N} = \int d \theta \exp(\beta D \cos^2(\theta-\alpha))$ is the normalisation and is equal to
\begin{align}
\tilde{N} &= 2 \pi \exp(\beta D/2) I_0(\beta D/2)
\end{align} 
where $I_0(x)$ is the zeroth order modified Bessel function of the first kind.

\subsection*{Average over disorder}
Since $\alpha_i$ are i.i.d's chosen from a distribution $p(\alpha)$, the strong law of large numbers implies that as $N \rightarrow \infty$, Eq. \ref{lambdainf} becomes
\begin{equation}
\Lambda(y_1,y_2) =\int_{0}^{2 \pi} d \alpha ~ p(\alpha) \log S
\end{equation}
We see that since the limit $N \rightarrow \infty$ is taken, with probability $1$, $\Lambda$ is the same for all disorder realizations and depends only on the distribution $p(\alpha)$.

To evaluate $S$, we define $z=\exp(i \theta)$ and convert the integral in Eq. \ref{S} into a contour integral over $z$ around a unit circle. We evaluate the integral via a Laurent series expansion of the integrand (see Appendix A). The result is:
\begin{align}
S(y_1,y_2) &= I_0(r)+ 2 \sum_{j=1}^{\infty} \frac{I_j(\beta D/2)}{I_0(\beta D/2)} I_{2j}(r) \cos 2j(\phi-\alpha)
\label{contourI}
\end{align}
where  $r=\sqrt{y_1^2+y_2^2}$ is the absolute value of the magnetisation and $\phi=\tan^{-1}(y_2/y_1)$  is its orientation. The $I_j$ represents the $j^{th}$ modified Bessel function of the first kind. .

Let $(y_1^*,y_2^*)$ extremise the r.h.s of Eq. \ref{rfR}. Both $y_1^*$ and $y_2^*$ are functions of $x_1$ and $x_2$, given by the solutions of the equations:
\begin{align}
x_{1,2} &= \frac{\partial \Lambda(y_1,y_2)}{\partial y_{1,2}}
\end{align}
The rate function $I(x_1,x_2)$ can then be written as
\begin{align}
I(x_1,x_2) = g(x_1,x_2)- \inf_{x_1,x_2}g(x_1,x_2)
\label{fullI}
\end{align}
where
\begin{equation}
g(x_1,x_2) = x_1 y_1^* +x_2 y_2^* -\Lambda(y
_1^*,y_2^*)-\frac{\beta (x_1^2+x_2^2)}{2}
\end{equation}
In the thermodynamic limit, the probability $P_{H,\beta}(C_N:x_1,x_2)$  in  Eq. \ref{ldp} is dominated by the minimum of $I(x_1,x_2)$, where $\frac{\partial I}{\partial x_1}=0$ and $\frac{\partial I}{\partial x_2}=0$, which yields $y_1^* = \beta x_1$ and $y_2^*=\beta x_2$. Note that the rate function is like a generalized free energy functional in that its minimum $\frac{1}{\beta} \inf_{x_1,x_2} I(x_1,x_2)$ provides the free energy of the system. By susbtituting $y_1^*$ and $y_2^*$ in Eq. \ref{fullI} we get
\begin{align}
I(x_1,x_2) &= \frac{\beta r^2}{2}- \log I_0(\beta r)
- \int_0^{2 \pi} d \alpha p(\alpha)\nonumber\\ &\log\left(1 + \sum_{k=1}^{\infty} 2 c_k \cos(2 k (\theta-\alpha)) \frac{I_{2k} (\beta r)}{I_0(\beta r)}\right)
\label{qrf}
\end{align}
where $r=\sqrt{x_1^2+x_2^2}$, $\theta = \tan^{-1}(x_2/x_1)$  and
\begin{equation} 
c_k =\frac{I_{k}(\beta D/2)}{I_0(\beta D/2)}
\label{ck}
\end{equation} 
 
Equation \ref{qrf} is the general expression of the free energy functional for the RCXY model on a fully connected graph for an arbitrary distribution of disorder.  The free energy of the system is equal to $\frac{1}{\beta} \inf_{x_1,x_2} I(x_1,x_2)$. Here $x_1$ and $x_2$ are the magnitudes of magnetisation in the $x$ and $y$ directions respectively and are the two order parameters of the system. Eq. \ref{qrf} is the main equation that we use to study different disorder distributions in the sections that follow.

We recover the pure XY model by setting $D=0$, in which case Eq. \ref{qrf} reduces to
\begin{equation}
I(x_1,x_2) = \frac{\beta r^2}{2} -\log I_0(\beta r)
\label{purerf}
\end{equation} 
which is isotropic in $x_1$ and $x_2$ and a function of $r$, agreeing with \cite{kirkpatrick}. The self-consistent equation 
for the magnetisation $r$ is:
\begin{equation}
\beta r = \beta \frac{I_1(\beta r)}{I_0(\beta r)}
\end{equation}
The system has a continuous transition as can be seen by expanding  RHS in powers of $r$ upto third order. We get 
\begin{equation}
\beta r \approx \frac{\beta ^2 r}{2} - \frac{\beta^4 r^3}{16}
\end{equation}
The transition from XY paramagnetic state ($r=0$) to a magnetic state ($r \neq 0$) occurs at $\beta_c =2$ ($T=1/2$). The magnetisation grows as $r \sim \sqrt{\beta-\beta_c}$, close to $\beta_c$.

For nonzero $D$, the phase diagram depends on the distribution of the disorder, given by $p(\alpha)$. For continuous transitions, the coefficient of the second order term in Eq. \ref{qrf} decides the location of the transition. To second order, we find 
\begin{align}
I(x_1,x_2) &\approx \frac{\beta}{4}(2-\beta) (x_1^2+x_2^2)-\frac{\beta^2 c_1}{2} x_1 x_2 <\sin 2 \alpha>\nonumber\\ &-\frac{\beta^2 c_1}{4} (x_1^2-x_2^2) <\cos 2\alpha>
\label{gct}
\end{align}
where $<>$ represents an average with respect to $p(\alpha)$. We observe that for distributions with $<\exp(2 i \alpha)>=0$, if there is a continuous transition, it is at $\beta_c=2$ independent of the value of $D$.  This holds for a large class of distributions, in particular for quadriperiodic distributions defined through $p(\alpha)=p(\pi/2+\alpha)$.

In the next two sections we study the phase diagram of the RCXY model for uniform and bimodal distributions of the crystal field disorder. 

\section{Uniform distribution}
\label{flatd1}

The uniform distribution of the anisotropy axis corresponds to
\begin{equation}
p(\alpha)=\frac{1}{2 \pi}~\forall~\alpha
\end{equation}
Substituting in Eq. \ref{qrf} , the rate function becomes
\begin{align}
I(x_1,x_2) &= \frac{\beta r^2}{2}- \log I_0(\beta r)-\frac{1}{2 \pi}  \int_0^{2 \pi} d \alpha \nonumber\\
& \log\left(1 + \sum_{k=1}^{\infty} 2 c_k \cos(2 k (\theta-\alpha)) \frac{I_{2k} (\beta r)}{I_0(\beta r)}\right)
\label{frf}
\end{align}
The integral over the disorder distribution can be performed exactly when $D \rightarrow \infty$ and also at large but finite $D$. We first study these two cases and then examine the case 
of arbitrary $D$ by expanding the integrand in powers of $r$. 

\subsection{Infinite D}
\label{flatD1}
The limit $D \rightarrow \infty$ forces each spin $s_i$ to point along or opposite to $\alpha_i$, thus reducing it to an Ising spin along the anisotropy axis.

As $D \rightarrow \infty$ the coefficients $c_k \rightarrow 1$. Setting $c_k =1$ $\forall~k$, we get
\begin{align}
I(x_1,x_2) &= \frac{\beta r^2}{2}- \log I_0(\beta r)-\frac{1}{2 \pi}  \int_0^{2 \pi} d \alpha \nonumber\\
&\log\left(1 + \sum_{k=1}^{\infty} 2 \cos(2 k (\theta-\alpha)) \frac{I_{2k} (\beta r)}{I_0(\beta r)}\right)
\end{align}
The summation inside the $\log$ term can then be done exactly using the identity \cite{abramowitz} : 
\begin{equation}
\sum_{k=1}^{\infty} \cos(2 k t) I_{2k} (x) = \frac{1}{2} (\cosh(x \cos t)-I_0(x))
\label{gfbessel}
\end{equation}
leading to
\begin{equation} 
I(r)=  \frac{\beta r^2}{2}- \frac{1}{2 \pi}  \int_0^{2 \pi} d \alpha \log(\cosh(\beta r \cos \alpha))
\label{infDI}
\end{equation}
The minimum of $I(r)$ w.r.t. magnetisation $r$ results 
in a self-consistent equation for $r$, given by 
\begin{equation}
r = \frac{1}{2 \pi} \int_0^{2 \pi} d\alpha \cos(\alpha) \tanh(\beta r \cos\alpha)
\label{qinfd}
\end{equation}
To find $\beta_c$, we expand $I(r)$ in Eq. \ref{infDI} in powers of $r$ till the fourth order:
\begin{equation}
I(r) = \frac{\beta r^2}{2}-\frac{\beta^2 r^2}{4}+\frac{\beta^4 r^4}{32}
\end{equation}
Since the coefficient of the $r^4$ term is positive, $\beta_c$ for the transition from XY ferromagnetic state to a paramagnet is found by equating the coefficient of $r^2$ to zero. This yields $\beta_c=2$, the same value as for the pure XY model.

The resulting model maps to a quenched random bond Ising model with correlated variables \cite{lubensky}, allowing a solution for the fully connected graph \cite{dv}. The self-consistent equation for magnetisation obtained above (Eq. \ref{qinfd}) agrees with the expression obtained in \cite{dv}.

Let us examine the low temperature behavior of the system. For $T=0$ the function $\tanh(\beta r \cos(\alpha)) =1$ if $\cos(\alpha)>0$ and $=-1$ if $\cos(\alpha)<0$. Hence in this case the magnetisation at $T=0$ is
\begin{equation}
r_0 =\frac{2}{\pi}
\end{equation}
 For nonzero low temperature, we use $\tanh z \approx \pm (1 - 2 \exp(-2 |z|)$ to obtain
\begin{equation}
r =\frac{2}{\pi}-\frac{\pi T^2}{4}
\end{equation}
Since the second term in Eq. \ref{infDI} is a function of $\beta r$, the internal energy for this model is proportional to $r^2$.  This implies that specific heat $C_v \sim T$ for low temperatures, vanishing as $T \rightarrow 0$. 

\subsection{Large D}
\label{flatD2}

To study the large $D$ behaviour, we employ the asymptotic expansion of $c_k$ \cite{abramowitz} in Eq. \ref{frf}.
\begin{equation}
c_k =\frac{I_k(\beta D/2)}{I_0(\beta D/2)} \approx 1-\frac{4 k^2}{8 \beta D+1}
\label{ckD}
\end{equation}
Differentiating Eq. \ref{gfbessel} twice we obtain the identity :
\begin{align}
& -\sum_{k=1}^{\infty} 4 k^2 \cos(2 k t)I_{2k}(x)\nonumber\\
 &= \frac{1}{2}(x^2 \cosh(x \cos t) sin^2 t-x \cos t \sinh(x \cos t))
\label{dgfbessel}
\end{align}
Using Eqs. \ref{gfbessel} and \ref{dgfbessel} and retaining terms only of order $1/D$, the rate function becomes
\begin{align}
I(r) &=  \frac{\beta r^2}{2}- \frac{1}{2 \pi}  \int_0^{2 \pi} d \alpha \log \left(\cosh(\beta r cos(\alpha))\right)\nonumber\\
&- \frac{\beta r^2}{16 \pi D} \int_0^{2 \pi} d \alpha \sin^2(\alpha)\nonumber\\ &+\frac{r}{16 \pi D} \int_0^{2 \pi} d \alpha \cos(\alpha) \tanh(\beta r \cos(\alpha))
\end{align}
For low $T$, the free energy functional $\phi(r)=\frac{1}{\beta} I(r)$, to leading order in $T$ is given by
\begin{equation}
\phi(r) = \frac{r^2}{2}-\frac{2}{\pi} r-\frac{r^2}{16D}+\frac{Tr}{4 \pi D}
\label{largedop}
\end{equation}
Equating $\partial \phi/\partial r =0$, we get the equation for magnetisation $r$ as
\begin{equation}
r-\frac{2}{\pi}-\frac{r}{8D}+\frac{T}{4 \pi D}=0
\end{equation}
For $T=0$, we find
\begin{equation}
r = \frac{2}{\pi} \left(1+\frac{1}{8 D}\right)
\label{maglarged}
\end{equation}
The increase proportional to $1/D$ from the $D \rightarrow \infty$ value is consistent with the $T=0$ mean field result of \cite{callen}.

For low finite temperatures, the leading order correction to $T=0$ value of $r$ is proportional to $T$ and is given by $r =\frac{2}{\pi} \left(1+\frac{1}{8 D}-\frac{T}{8 D}\right)$.

Since the internal energy $U$ is proportional to $r^2$, it is 
linear in $T$, implying that the specific heat $C$ goes to a constant as $T$ approaches zero for large finite $D$. This is because for $T<<D$ the spins make excursions of low amplitude $\delta s_i$ around their ground state positions, with $\langle \delta s_i^2 \rangle << T/D$. This ``Dulong-Petit'' contribution results in a finite value of $C$. When $D = \infty$, these excitations are forbidden, leading to $C \rightarrow 0$ as $T \rightarrow 0$. The energy spectrum develops a gap for $D= \infty$ and goes to zero continuously for all finite values of $D$.

\subsection{Expansion in powers of $r$ for finite $D$ }
\label{flatD3}
\begin{figure}
     \begin{subfigure}[b]{0.45\textwidth}
         \includegraphics[width=\textwidth]{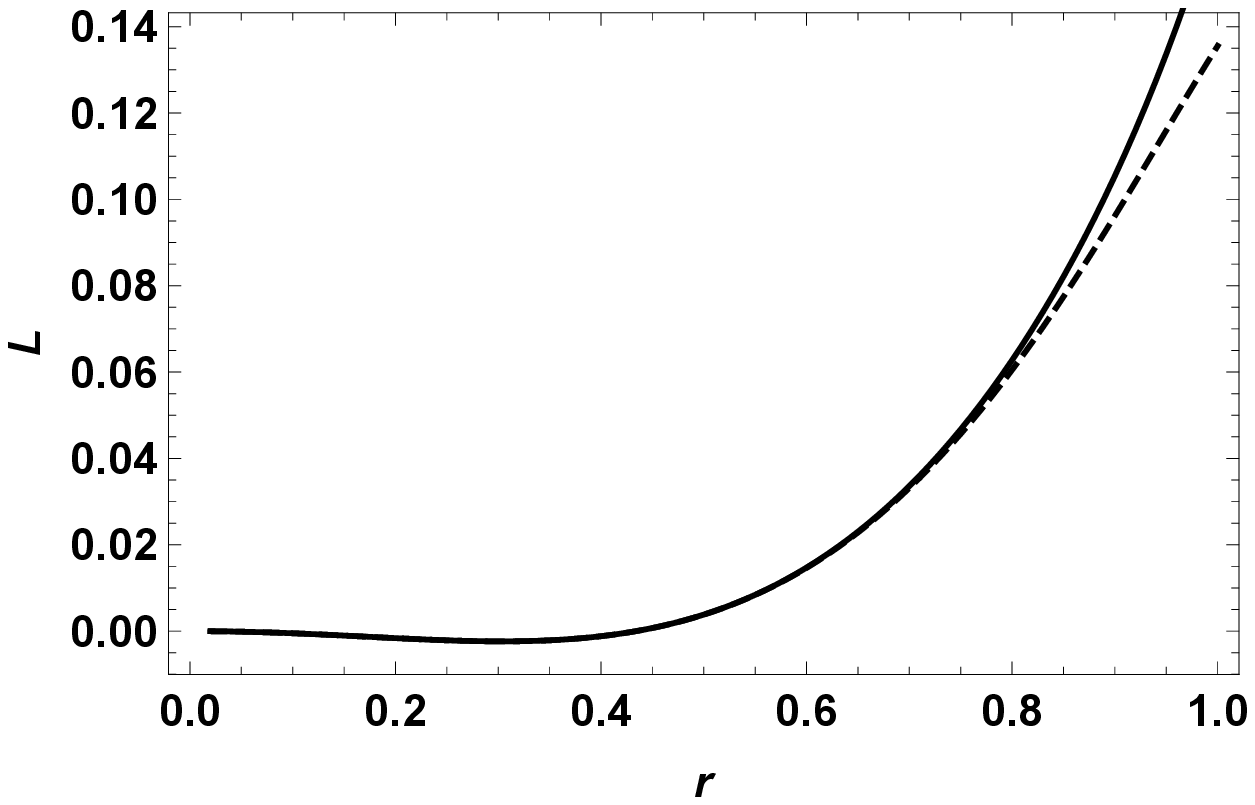}
         \caption{}
     \end{subfigure}
     \hfill
     \begin{subfigure}[b]{0.45\textwidth}
         \includegraphics[width=\textwidth]{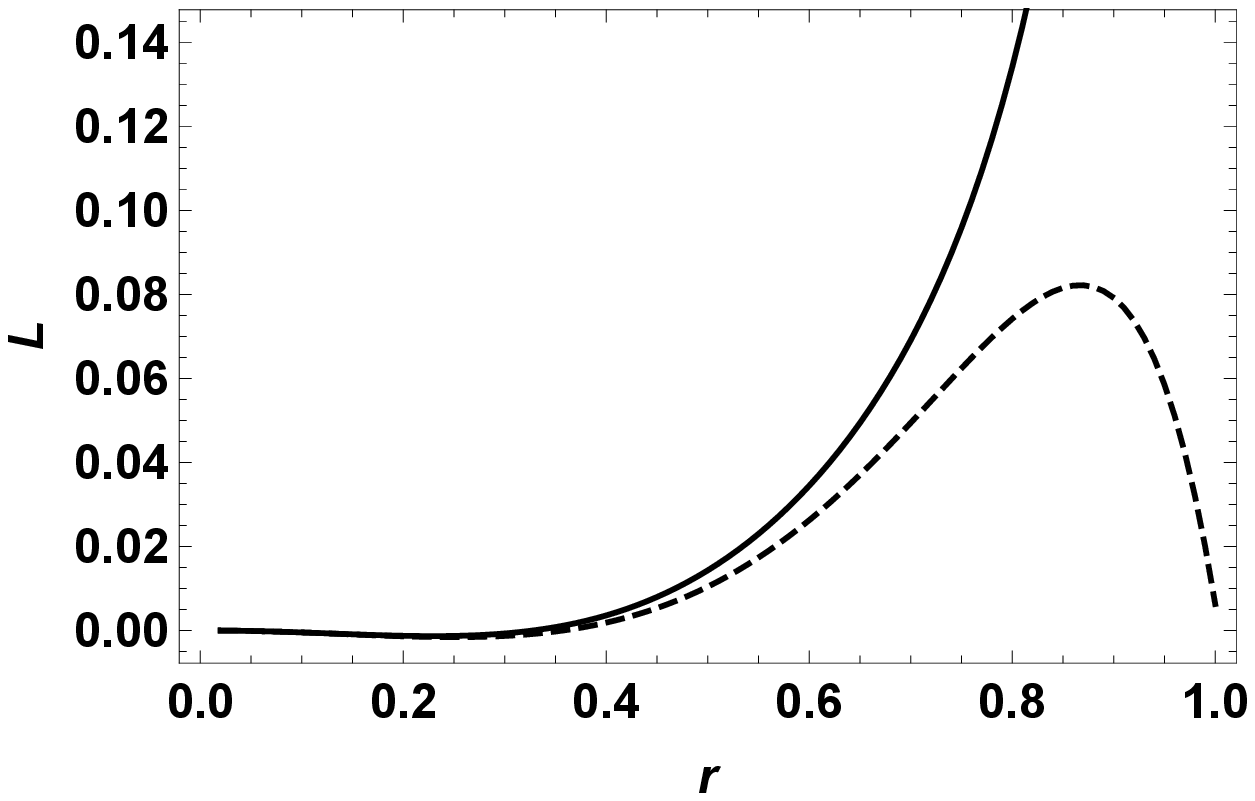}
         \caption{}
     \end{subfigure}
     \hfill

             \caption{Truncated Landau functional for uniform distribution obtained by expanding till $10^{th}$ ($12^{th}$) power in $r$ shown as the dotted (solid) line, for a) $D=0$ and $\beta=2.1$; b) $D=2$ and $\beta=2.1$. }
             \label{flatd}
             \end{figure}
When the LHS of Eq. \ref{frf} is expanded in powers of $r$, the integration over  $\alpha$ eliminates terms which are not isotropic in $x_1$ and $x_2$ and only the terms that are functions of $r=\sqrt{x_1^2+x_2^2}$ survive. Thus $I(x_1,x_2)$ is a function of $r$ alone for all values of $D$, for uniform  distribution of $\alpha$. For example, the expansion to $8th$ order, reads
\begin{align}
L(r)&= \frac{\beta r^2 (2-\beta)}{4}+\frac{(\beta r)^4}{64}(1+c_1^2)-\frac{(\beta r)^6}{576}(1+3c_1^2)\nonumber\\ +&
\frac{(\beta r)^8}{2}\left(\frac{11}{24576} 
+ \frac{49 c_1^2}{18432} + \frac{c_2^2}{73728}+\frac{3 c_1^4}{4096}-\frac{6 c_1^2 c_2}{3689}\right)\nonumber\\-&...
\label{flf}
\end{align}
In analogy with Landau theory, we have denoted the power series expansion of $I(r)$ by $L(r)$.
We observe that the terms in the expansion alternate in sign for all values of $D$. Close to the transition temperature, $r$ is small and it suffices to keep second and fourth order terms. Since the latter is always positive, we locate the critical point by equating the coefficient of second order term (which is independent of $D$) to zero. This gives 
\begin{equation}
\beta_c =2 ~~\forall~ D
\end{equation}
The limit $D=0$, may be recovered on noting that the alternating series can be summed and is equal to $\log I_0(\beta r)$ (see Eq. \ref{purerf}). The coefficients $a_i$ , associated with $i^{th}$ power of $r$, decrease monotonically with $i$ and the series converges. But if $D$ is nonzero, the montonicity of the coefficients is not retained. Their magnitude increases beyond a certain value of $i$ which depends on $D$. We tabulate the coefficients upto $i=12$ in Table \ref{tab:01} for $D=0, 1, 10$ and $1000$ at $\beta=2$ to illustrate this.
\begin{table}[H]
\begin{center}
\begin{tabular}{|c|c|c|c|c|c|}
\hline
$D$ & $a_4$ & $a_6$ & $a_8$ & $a_{10}$ & $a_{12}$ \\
\hline
0 & 0.25 & -0.1111 & 0.0573 & -0.0317 & 0.01825\\\hline
1 & 0.3717 & -0.2734 & 0.2423 & -0.2361 & 0.2443\\\hline
10 & 0.4875 & -0.4277 & 0.4487 & -0.5169 & 0.6321\\\hline
1000 & 0.4998 & -0.4443 & 0.4719 & -0.5507 & 0.6819\\\hline
\end{tabular}
\end{center}
\caption{Coefficients of $r^n$ for different values of $D$ at $\beta =2$ in Eq. \ref{flf}.}
\label{tab:01}
\end{table}
Due to poor convergence for $D \neq 0$, the series cannot be used to study the low temperature behaviour. The behaviour of the free energy functional changes for large $r$, depends on the term at which we truncate the expansion. Figure \ref{flatd} shows the free energy to  $10^{th}$ and $12^{th}$ order for $D=0$ and $D=2$ for $\beta =2.1$. The possibility of a first order transition at low temperatures cannot be completely ruled out, but in our investigation till order $24$ , we did not find any evidence of it. 
\section{Bimodal distribution}
\label{bimodald}
Now consider the distribution 
\begin{equation}
p(\alpha)= p \delta(\alpha-0)+(1-p) \delta (\alpha-\pi/2)
\label{bpd}
\end{equation}
i.e, a fraction $p$ of the spins experience a crystal field pointing along the $x$-axis, while the remaining fraction $(1-p)$ are in a crystal field along the $y$-axis. The cases $p=0$ and $p=1$ correspond to no disorder.

Substituting in Eq. \ref{qrf}, the rate function is
\begin{align}
I(x_1,x_2) &= \frac{\beta r^2}{2}- \log I_0(\beta r)\nonumber\\ -& p \log\left(1 + \sum_{k=1}^{\infty} 2 c_k \cos(2 k \theta) \frac{I_{2k} (\beta r)}{I_0(\beta r)}\right)\nonumber\\
  -&(1-p) \log\left(1 + \sum_{k=1}^{\infty} 2 c_k \cos(k (\pi- 2  \theta)) \frac{I_{2k} (\beta r)}{I_0(\beta r)}\right)
\label{blf}
\end{align}
where again $r=\sqrt{x_1^2+x_2^2}$, $\theta =\tan^{-1}x_2/x_1$ and $c_k =I_{k}(\beta D/2)/I_0(\beta D/2)$. The minimum of 
this function for a given set of parameters $p, \beta$ and $D$ gives the free energy of the model. 

We first discuss the phase diagram of the model in the limit of  $D \rightarrow \infty$, in which case the summations inside the log term can be performed.
 \subsection{D = $\bf{\infty}$}
\label{infdbm}
We obtain the rate function by making use of Eq. \ref{gfbessel} in Eq. \ref{blf}, with $c_k=1$. We get
\begin{align}
I(x_1,x_2) &= \frac{\beta (x_1^2+x_2^2)}{2}- p \log(\cosh(\beta x_1))\nonumber\\ &-(1-p) \log(\cosh(\beta x_2))
\end{align}
Minimising the rate function gives two self-consistent equations for the order parameters in $x$ and $y$ directions as
\begin{eqnarray}
x_1 &=& p \tanh \beta x_1\\
x_2 &=& (1-p) \tanh \beta x_2
\label{opbd}
\end{eqnarray}
For $0 \le p \le 1$, the RCXY model reduces to two uncoupled Ising models, with a fraction $p$ of spins along the $x$-direction ($x_1 =p$) and fraction $1-p$ of spins aligned along the $y$-direction ($x_2=1-p$) in the ground state. 
\begin{figure}
\includegraphics[scale=0.7]{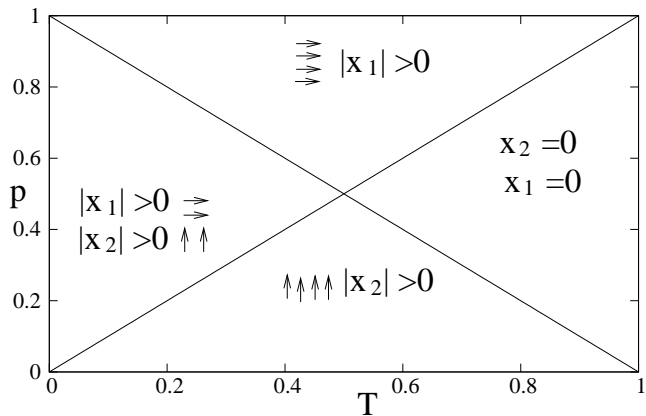}
\caption{$p-T$ phase diagram for bimodal distribution for infinite D.}
\label{bidinf}
\end{figure}
In the  $p-T$ plane, there are two lines of continuous transitions, one with $T=p$ (separating $x_1=0$ from $x_1 \neq 0$) and the other with $T=(1-p)$ (separating $x_2=0$ from $x_2 \neq 0$). The phase diagram has four phases as shown in Fig. \ref{bidinf}. These phases are separated by four critical lines, all of which lie in the mean field Ising universality class. These lines intersect at $(T,p)=(\frac{1}{2},\frac{1}{2})$. For $T<1/2$ the phase between the critical lines $p=T$ and $1-p=T$ is a mixed phase with $x_1 \neq 0$ and $x_2 \neq 0$, with. a total lack of coupling.


\subsection{Finite D}
For arbitrary $D$ near the critical loci, we expand $I(x_1,x_2)$ in Eq. \ref{blf} in powers of $x_1$ and $x_2$ as they are small. This then gives us the Landau free energy expansion of the functional with known coefficients.

The lowest order term in expansion of $I_{2 k}(\beta r)/I_0(\beta r)$ is of order $r^{2 k}$. Hence the expression  
  $\cos(2 k \theta) \frac{I_{2 k}(\beta r)}{I_0(\beta r)}$, has terms of order higher than four for $k>2$. We expand Eq. \ref{blf}, by keeping terms only till $k=2$. 

The result is a two parameter Landau functional of the form
\begin{align}
L(x_1,x_2) &= a_+ (x_1^2+x_2^2) + a_- (x_1^2-x_2^2)+u_1 x_1^4\nonumber\\ &+u_2 x_2^4+2 u_{12} x_1^2 x_2^2
\label{lfbd}
\end{align}
We denote this function by $L(x_1,x_2)$ to distinguish it from the full rate function $I(x_1,x_2)$. Here, 
\begin{align}
a_+ &= \frac{\beta}{4} (2-\beta)\nonumber\\a_- &=(1-2p) \frac{\beta^2 c_1}{4}\nonumber\\u_1 &=\frac{\beta^4}{192} (3-c_2+6 c_1^2+8 (2p-1)c_1)\nonumber\\u_2 &=\frac{ \beta^4}{192} (3-c_2+6 c_1^2-8 (2p-1)c_1)\nonumber\\u_{12} &=\frac{\beta^4}{64} (1+c_2-2 c_1^2)
\label{apam}
\end{align}

The phase diagram resulting from this functional is worked out in detail in Appendix B; it depends on the value of the ratio $s$, defined as $s= \frac{u_1 u_2}{u_{12}^2}$. Here we merely summarize the results. There are four possible states: $(0,0)$, $(0,x_2)$, $(x_1,0)$ and $(x_1,x_2)$.  For $s \le 1$, the phase $(x_1,x_2)$ is not stable and the system exhibits two curves of continuous transitions given by the  equations $a_+=a_-$ and $a_+ =- a_-$. These two meet at the point $(a_+,a_-)=(0,0)$ in the $(a_+,a_-)$ plane. This point is a bicritical point. It is also an end point of a first order spin flop line separating the two Ising ordered phases with finite magnetisations in the $x$ and $y$ directions respectively (transverse and longitudinal Ising phases respectively). For $s>1$, all four phases are possible and the phase diagram now has four critical curves meeting at $(a_+,a_-)=(0,0)$. This point is now a  tetra-critical point. 

We now use these results to obtain the phase diagram of the bimodal RCXY defined by Eq. \ref{blf}, as a function of $D,T$ and $p$.
\subsubsection{Pure Case(p=0)} 
For $D=\infty$ there is a transition to the longitudinal Ising phase at $T=1$, as discussed in Section \ref{infdbm}. For finite $D$ the coefficient of the $x_1^2 x_2^2$ term in Eq. \ref{lfbd} is not zero ; $x_1$ and $x_2$ are coupled to each other in general.

The ratio $s=\frac{u_1 u_2}{u_{12}^2}$ in this case is $1$ for $D=0$ and decreases with increasing $D$. Hence the mixed phase $(x_1,x_2)$ is not stable and the system has a bicritical point where the two critical curves meet. These two critical curves are given by $a_+=a_-$ and $a_+=-a_-$. They separate the paramagnetic phase from the Ising phase aligned longitudinally ($(0,x_2)$) and transversely($(x_1,0)$) respectively. The equations of the critical curves are 
\begin{eqnarray}
2-\beta_c =  \pm \beta_c c_{1,c}
\end{eqnarray}
where $\beta_c=1/T_c$ and $c_{1,c} =I_1(\beta_c D/2)/I_0(\beta_c D/2)$. 

\begin{figure}
\includegraphics[scale=0.7]{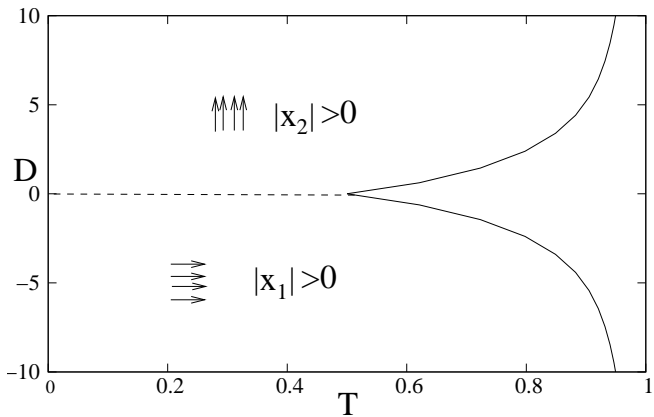}
\caption{Phase diagram for the pure XY model with crystal-field in the $y$-direction. The solid lines are critical curves separating the Ising phases from paramagnetic phase. across the dotted line there is a first order spin flop transition between the two Ising phases. As $D \rightarrow \pm \infty$, the two critical lines approach $T=1$.}
\label{purepd}
\end{figure}

The critical curves $T_c=\frac{1\pm c_{1,c}}{2}$ are shown in the phase diagram, in Fig.\ref{purepd}. They separate the paramagnetic state from a state with longitudinal (transverse) order for $D>0$ $(D<0)$. There is a first order spin flop transition on crossing the locus $T<1/2$, $D=0$, from  transverse to a longitudinal phase. The locus terminates in a bicritical point at $T=1/2$, $D=0$.

\subsubsection{Quadriperiodic bimodal disorder distribution ($p=1/2$)}
We study the case $p=1/2$ first. In this case, $a_-=0$ and $u_1=u_2$. Also $s=(u_1u_2)/u_{12}^2$ is greater than $1$ for all values of the crystal field strength $D \neq 0$. The Landau functional in this case becomes symmetric in $x_1$ and $x_2$ and takes the form
\begin{equation}
\label{lfes}
L(x_1,x_2) = a_+ (x_1^2+x_2^2) +u_1 x_1^4+u_1 x_2^4+2 u_{12} x_1^2 x_2^2
\end{equation}
There is only one line of continuous transitions, given by equating $a_+$ to $0$. This gives $\beta_c=2$ $\forall D$. This line of continuous transition separates the XY ferromagnetic phase from a paramagnet. Hence the phase boundary in this case is the same as for the uniform distribution. However the ordered phase is different. It is now a four-fold degenerate phase with $|x_1|=|x_2|$.

\subsubsection{Asymmetric bimodal distribution, $0<p<1$}
In this case there is a crystal field pointing in the $x$-direction for a randomly chosen fraction $p$ of the spins and in the $y$-direction for the remaining fraction $1-p$. The effect of disorder is maximum 
for $p=1/2$. The ratio $s=\frac{u_1 u_2}{u_{12}^2}$ is a function of $p$ and $w=\beta D$ alone. 

For $D=0$ and hence for $w=0$, the ratio $s=1$. For a fixed $w$, $s > 1$ for $p_{l}(w) < p < p_{u}(w)$, where $p_{l}(w)$ and $p_{u}(w)$ are functions of $w$ alone which rapidly approach $0$ and $1$ respectively as $w$ increases (see Table \ref{tab:2}). For $p<p_l(w)$ and $p>p_u(w)$, there is no mixed phase for any value of $p$ and $T$.
\begin{table}[H]
\begin{center}
\begin{tabular}{|c|c|c|}
\hline
$w$ & $p_l(w)$ & $p_u(w)$  \\
\hline
0.1 &0.01582 & 0.98418 \\\hline
0.5 & 0.01445 & 0.98555 \\\hline
1.0 & 0.01107 & 0.98893 \\\hline
1.5 & 0.00739 & 0.99261 \\\hline
2.0& 0.00449 & 0.99551\\\hline
3.0 & 0.00144 & 0.99856\\\hline
\end{tabular}
\end{center}
\caption{Lower and upper threshold on probability $p$ such that for $p<p_l(w)$ and $p>p_u(w)$ for a given $w$, there is no mixed phase}
\label{tab:2}
\end{table}
In the next two subsections we study the phase diagram for a fixed $D$ and fixed $w$ separately . The phase diagram consists of four critical curves in the $p-T$ plane, meeting at a tetra-critical point. 

The two critical curves separating the $(0,0)$ and $(0,x_2)$ phases 
and $(0,0)$ and $(x_1,0)$ phases are given by the equations $a_+=a_-$ and $a_+=-a_-$ respectively. Substituting for $a_+$ and $a_-$ as in Eq. \ref{apam}, we get
\begin{eqnarray}
T_c=\frac{(1 \pm (1-2p_c) c_{1,c})}{2}
\label{isingp1}
\end{eqnarray}
as the equations of the two critical curves, separating the transverse Ising and longitudinal Ising phases from the paramagnetic phase. Here again $c_{1,c} =I_1(\beta_c D/2)/I_0(\beta_c D/2)$.

Two other critical curves separate the $(x_1,0)$ and $(0,x_2)$ phases from the mixed phase, represented as $(x_1,x_2)$. They are given by 
$a_+ = \alpha_1 a_-$ and $a_+ = -\alpha_2 a_-$ respectively. These 
two conditions give the equations of critical curves to be
\begin{eqnarray}
T_c = \frac{(1 \mp (1-2p_c) \alpha c_{1,c})}{2}
\label{mixedp1}
\end{eqnarray}
where $\alpha=\alpha_1= \frac{u_1+u_{12}}{u_1-u_{12}}$ and $\alpha=\alpha_2= \frac{u_2+u_{12}}{u_2-u_{12}}$ respectively as defined in the Appendix B. Note that $u_{12},u_1$ and $u_2$ are also functions of $p_c, D$ and $T_c$.
\subsubsection{Phase diagram with fixed $D$}
\begin{figure}
     \centering
     \begin{subfigure}[b]{0.45\textwidth}
         \centering
         \includegraphics[width=\textwidth]{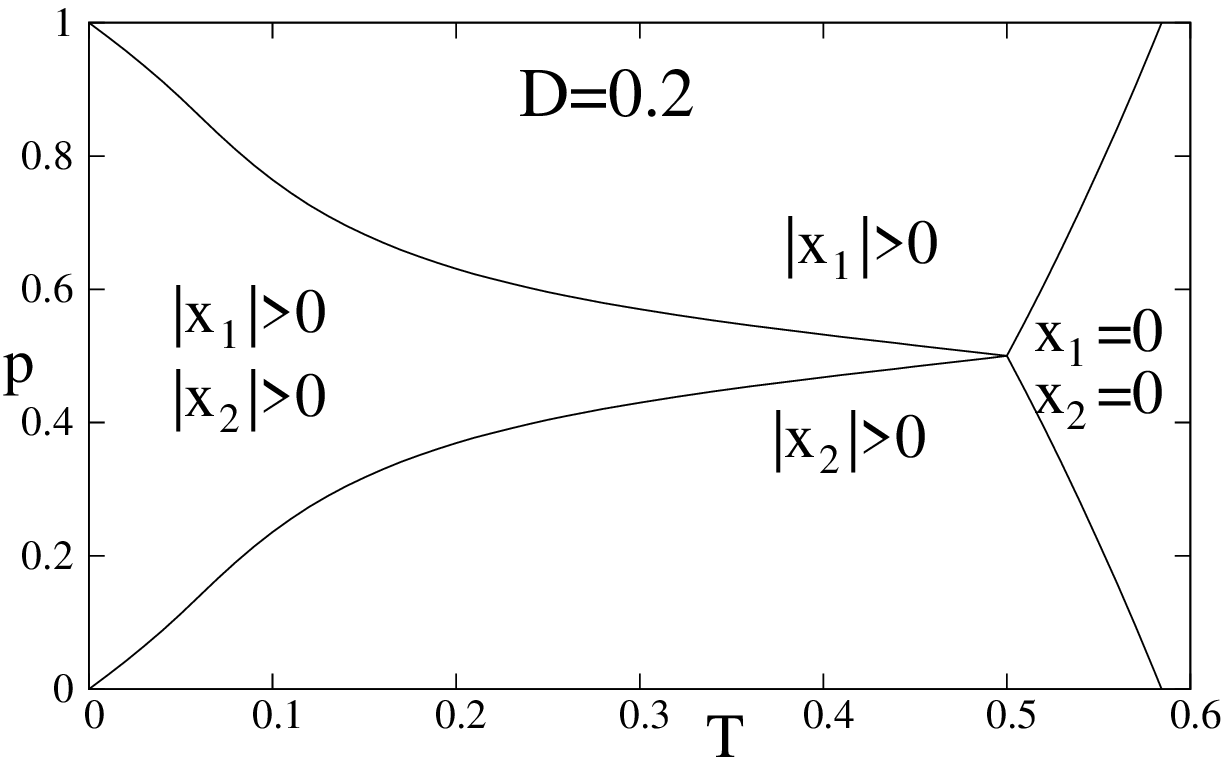}
         \caption{\footnotesize{}}
         \label{D0p2}
     \end{subfigure}
     \hfill
     \begin{subfigure}[b]{0.45\textwidth}
         \centering
         \includegraphics[width=\textwidth]{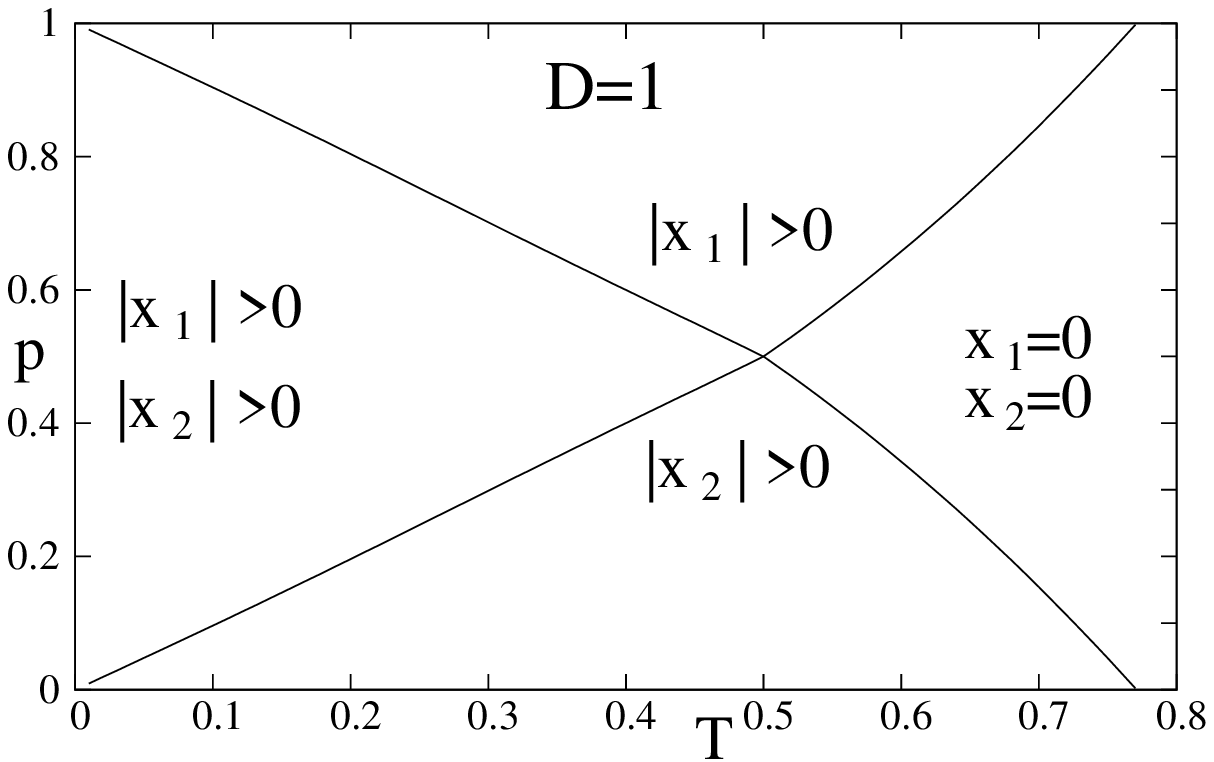}
         \caption{\scriptsize{}}
         \label{D1}
     \end{subfigure}
     \hfill
             \caption{Phase diagram in the $p-T$ plane for a)$D=0.2$ and b) $D=1$ for the bimodal distribution. There is a tetra-critical point at $(1/2,1/2)$ from which four critical curves emanate, separating the four phases.}
        \label{fixedDpd}
\end{figure}

For any finite $D$, as $\beta \rightarrow \infty$, $w \rightarrow \infty$, there is a mixed phase for all values of $p$ at $T=0$. The phase diagram has a tetra-critical point at $T=1/2$ and $p=1/2$, where the four critical curves given by Eqs. \ref{isingp1} and \ref{mixedp1} meet. The phase diagram for $D=0.2$ and $D=1$ in the $p-T$ plane is plotted in Fig. \ref{fixedDpd}. As $D$ increases,  the area under the mixed phase increases and the phase diagram rapidly converges to the $D \rightarrow \infty$ phase diagram given in Fig. \ref{bidinf}. 
All the critical curves in the phase diagram belong to the mean field Ising universality class as does the tetra-critical point. The critical curves are straight lines only near the tetra-critical point and develop non-linearity at low temperatures, unlike the standard mean-field solutions \cite{chaikin}.

It is instructive to examine the magnetic susceptibilities corresponding to the two order parameters $x_1$ and $x_2$. We define $\chi_{11} =(\partial x_1/\partial h_1)_{h_1 \rightarrow 0}$ and $\chi_{22}= (\partial x_2/\partial h_2)_{h_2 \rightarrow 0}$ as the susceptibilities corresponding to the the magnetisations $x_1$ and $x_2$ respectively ($h_1$ and $h_2$ are the uniform external field in the directions $x$ and $y$). 

\begin{figure}
         \includegraphics[width=0.45\textwidth]{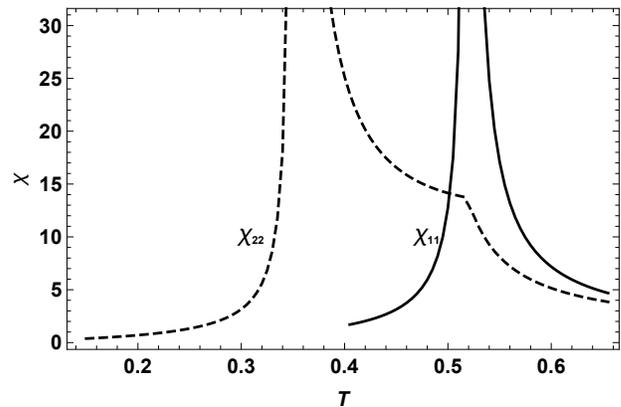}
             \caption{Transverse ($\chi_{11}$) and longitudinal ($\chi_{22}$) susceptibility in the case of asymmetric bimodal distribution with $D=0.2$ and $p=0.4$.}
             \label{d0p2p}
             \end{figure}

To study the singularities along the two different critical curves,  we plot $\chi_{11}$ and $\chi_{22}$ for $p=0.4$ for $D=0.2$ in Fig. \ref{d0p2p}. As expected $\chi_{11}$ diverges near the paramagnetic to Ising transition and $\chi_{22}$ diverges near the Ising to mixed phase transition. Interestingly, though $\chi_{22}$ does not diverge near the paramagnetic to Ising transition, it exhibits a discontinuity of slope. Similar behaviour is seen also for $p>0.5$, where the roles of $\chi_{11}$ and $\chi_{22}$ are interchanged.

\subsubsection{Phase diagram with fixed w}
The phase diagram in the $p-T$ plane for fixed $w$ is similar to the phase diagram for fixed $D$. The main difference is at $T=0$. For $T=0$, with $w$ finite, the mixed phase occurs only between two threshold values of $p$. 

Taking $T=0$ in Eq. \ref{mixedp1}, we find a lower threshold on $p$ through the self-consistent equation
\begin{equation}
p_{0l}(w) =\frac{1}{2}-\frac{1}{2 \alpha_1 c_1}
\label{p0l}
\end{equation}
where $p_{0l}(w)$ is the critical value of $p$, separating the mixed and the longitudinal Ising phases at $T=0$. Note that $\alpha_1=\frac{u_1+u_{12}}{u_1-u_{12}}$ appearing on r.h.s in Eq. \ref{p0l} is also a function of $p_{0l}(w)$. The analogous upper threshold is given by 
\begin{equation}
p_{0u}(w) =\frac{1}{2}+\frac{1}{2 \alpha_2 c_1}
\label{p0u}
\end{equation}
where $\alpha_2=  \frac{u_2+u_{12}}{u_2-u_{12}}$.
\begin{figure}
\includegraphics[scale=0.55]{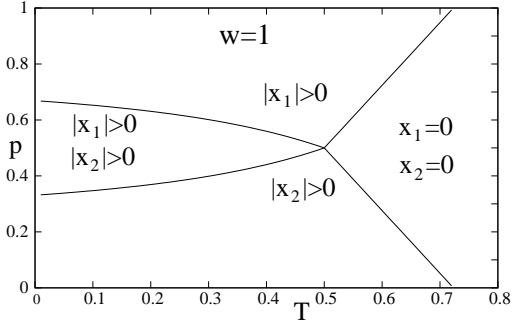}
\caption{Phase diagram for $w=1$ in the $p-T$ plane. Four critical curves meet at a tetra-critical point at $(1/2,1/2)$ for all $w$. The phases represented by $|x_1|>0$ and $|x_2|>0$ are transverse and longitudinal Ising phases with $|x_2|=0$ and $|x_1|=0$ respectively.}
\label{pTpd}
\end{figure}

It is instructive to plot the phase  diagram in the $(T,p)$ plane for fixed $w$ (Fig. \ref{pTpd}). Four critical curves meet at $p=1/2$ and $T=1/2$, which is thus a tetra-critical point. The critical curves are straight only near the tetra-critical point. The $y$-axis intercepts  
$p_{0l}(w)$ and $p_{0u}(w)$  of the two critical curves approach $0$ and $1$ repectively as $w \rightarrow \infty$.

As $p$ increases the Ising phases shrink and the critical curves approach each other. At $p=1/2$, the critical temperature becomes independent of $w$ and there is a single transition at $T=1/2$ for all values of $w$ from the disordered to the mixed phase ( $(0,0)$ to $(x_1,x_2)$). 

\subsection{Canted state at large D}
In this subsection, we address the nature of the mixed state, and show that the magnetization vectors are canted. At $T=0$, the exact results for $D=\infty$ in Sec. \ref{infdbm} give a ground state with a fraction $p$ of spins aligned along $x$ and a fraction $1-p$ of spins aligned along $y$. On the other hand if $D=0$, the ground state is rotationally invariant with $r=\sqrt{x_1^2+x_2^2} =1$. For finite $D$, we use a large $D$ expansion as in Sec. \ref{flatD2} for the uniform distribution. 
\begin{figure}
         \includegraphics[width=0.45\textwidth]{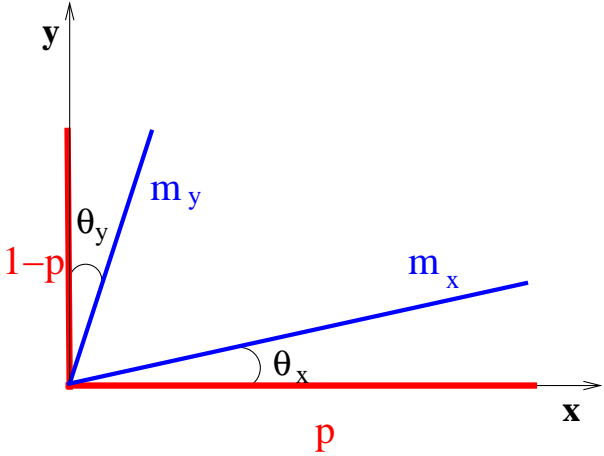}
             \caption{The two red vectors respresent the average  magnetisation vectors along $x$ and $y$ directions for $D=\infty$ at $T=0$, which have magnitude $p$ and $1-p$ respectively. Blue vectors represent the canted average magnetic vectors for large finite $D$ at $T=0$. We have taken $p$ such that $p > 1-p$ and hence $\theta_x < \theta_y$.}
             \label{cing}
             \end{figure}
Using Eq. \ref{ckD} for $c_k$ for large $D$ in Eq. \ref{blf} and taking $\beta$ to be large, the rate function reduces to 
\begin{align}
I(x_1,x_2) &= \frac{\beta}{2} (x_1^2+x_2^2) -p \beta x_1 -(1-p) \beta x_2\nonumber\\&-p\log \left(1+\frac{\beta^2 x_2^2}{1+8 \beta D}-\frac{\beta x_1}{1+8 \beta D}\right)\nonumber\\&-(1-p)\log \left(1+\frac{\beta^2 x_1^2}{1+8 \beta D}-\frac{\beta x_2}{1+8 \beta D}\right) 
\end{align}
Keeping terms till order $1/D$, we obtain 
\begin{align}
I(x_1,x_2) &= \frac{\beta}{2} (x_1^2+x_2^2)-p \beta x_1\nonumber\\&- (1-p) \beta x_2-\frac{(1-p) \beta}{8D} x_1^2-\frac{p \beta}{8 D} x_2^2
\end{align}
Equating partial derivatives w.r.t $x_1$ and $x_2$ to $0$, we obtain
\begin{align}
x_1 &= p \left(1-\frac{T}{8 D}+\frac{(1-p)}{4 D}\right) \nonumber\\x_2 &= (1-p) \left(1-\frac{T}{8 D}+\frac{p}{4 D}\right)
\label{bldc}
\end{align}
Equation \ref{bldc} describes a state in which the magnetisation vectors $\overrightarrow{m}_x$ and $\overrightarrow{m}_y$ are canted away from the $x$ and $y$ axes respectively as depicted in Fig. \ref{cing}, with canting angles $\theta_x$ and $\theta_y$ (which are small for large $D$). To leading order in $1/D$ we may write
\begin{align}
\overrightarrow{m}_x =m_x (\hat{x}+\theta_x \hat{y})~;~\overrightarrow{m}_y=m_y (\theta_y \hat{x}+\hat{y})
\end{align}
where $m_x=p$ and $m_y=1-p$ for $T=0$. Comparing this with Eq. \ref{bldc}, we get the canting angles at $T=0$ as 
\begin{align}
\theta_x=\frac{1-p}{4 D}~;~ \theta_y =\frac{p}{4 D}
\end{align}
While the crystal field $D$ tries to align the spins along the site with $x$ or $y$ axis depending on the value of $\alpha$, the mean field produced by other spins forces canting, and spin makes a small angle with the preferred axis. 

 The low $T$ phase for finite $D$ differs from that obtained with $D=\infty$. The specific heat shows the same behaviour with the uniform distribution : it approaches zero for $D = \infty$ and is constant for finite $D$ as $T \rightarrow 0$.

\section{Discussion}
\label{discuss}

We studied the RCXY model for different distributions of the disorder orientation. We found a remarkable constancy of $T_c$ for all distributions which satisfy $<\exp(2 i \alpha)>=0$, which includes quadriperiodic distributions for which $p(\alpha)= p(\alpha+\frac{\pi}{2})$. Uniform and symmetric bimodal distributions are examples of quadriperiodic distributions that we have studied in detail in this paper. In both cases, there is a single transition at $T_c=1/2$ from a mixed magnetic phase to a paramagnetic phase. The nature of the mixed phase depends on the distribution of disorder as can be seen by looking at the disorder-averaged ground state which inherits the symmetry of $p(\alpha)$. 

In the case of asymmetric bimodal distribution the asymmetry of the distribution results in a new ground state, namely the mixed phase in which the magnetisation is canted in two different directions for all finite values of the crystal field strength $D$. The ground state for $D=\infty$ is not canted, with spins aligned completely in the $x$ or $y$ direction. 

We find that in general the behaviour of RCXY for finite crystal field strength $D$ is different from the behaviour for $D = \infty$. The specific heat vanishes at $T = 0$ for $D=\infty$, but approaches a finite value for finite $D$. This is also reflected in the fact that $D=\infty$ RCXY can be mapped to correlated random bond Ising model \cite{dv}. We also extracted the disorder-averaged ground state of the model in the large $D$ limit and confirmed an earlier zero temperature mean field calculation where the order parameter at zero temperature was shown to decay as $1/D$ for uniform distribution \cite{callen}


Similar studies can be carried out for the random anisotropy model 
for vector spins with a number of components $m>2$. In particular the critical behvaiour can be studied easily by obtaining an expansion till fourth order in the order parameter $r$ for uniform distribution of the disorder. This yields the critical temperature for these models to be $1/m$, independent of the strength of the crystal field on a fully connected graph. However the full rate function needed to obtain the low temperature behaviour is non trivial due to the integrals involved in the calculation. 

We have recently studied the XY model on a fully connected graph in the presence of quenched random magnetic field (RFXY) drawn from different symmetric distributions \cite{sumedhabarma}. In that case, the disorder is in the field conjugate to the order parameter and has a much stronger effect. Not only $T_c$ but also the nature of the transition changes as a function of the strength of the magnetic field. The RFXY phase diagram consists of a line of second order transitions meeting a line of first order transitions at a tricritical point. Quenched random crystal field orientation disorder on the other hand does not couple directly with the order parameter and has a weaker effect. As we have seen, it does not change $T_c$ for any quadriperiodic distribution. It would be interesting to explore the quadriperiodic distribution of the random crystal field orientation on regular random graphs, in particular to see if $T_c$ stays unchanged.
\section{Acknowledgements} 
S acknowledges C. Gowdigere for discussions. M.B. acknowledges support under the DAE Homi Bhabha Chair Professorship of the Department of Atomic Energy, India. 

\appendix

\section{}

We solve the integral in Eq. \ref{S} using contour integration. The integral is
\begin{equation}
S =\frac{1}{\tilde{N}} \int_{0}^{2 \pi} \exp(\beta D \cos^2(\theta-\alpha)+x_1 \cos \theta +x_2 \sin \theta) d \theta
\end{equation}
where $\tilde N = \int_{0}^{2 \pi} \exp(\beta D \cos^2(\theta-\alpha_i))$.

We convert these integrals to contour integrals around a unit circle in the complex plane , by making a substitution $z =e^{i \theta}$ and $z_0 = e^{-i \alpha}$. Substituting, we get
\begin{align} 
S &=\frac{e^{\beta D/2}}{i \tilde N} \oint \frac{d z}{z} \exp \left(\frac{\beta D}{4} (z^2 z_0^2+z^{-2} z_0^{-2})\right)\nonumber\\ 
&\exp\left(\frac{x_1}{2} (z+z^{-1})+\frac{x_2}{2i}  (z-z^{-1})\right)
\label{App1}
\end{align}
 We define two new variables: $a=\frac{x_1-ix_2}{2}$ and $b= \frac{\beta D z_0^2}{4}$. The integrand in Eq. \ref{App1} has a form $f(z)/z$,  where $f(z)= \exp(b z^2+\bar{b} z^{-2}) \exp(a z+\bar{a}z^{-1})$. We can solve the integral using the Residue theorem. We get, $\tilde{N} S =2 \pi e^{\beta D/2} A_0$, where $A_0$ is the coefficient of the $z^0$ term in the expansion of $f(z)$. The function $f(z)$ can be expanded in terms of modified Bessel functions  of the first kind as follows
\begin{align}
& \exp(b z^2+\bar{b} z^{-2}) \exp(a z+\bar{a}z^{-1})\nonumber\\ &= \left(I_0(\beta D/2)+\sum_{j=1}^{\infty} ((zz_0)^{2 j}+(zz_0
)^{-2 j}) I_j(\beta D/2)\right)\\\nonumber
& \left(I_0(r)+\sum_{j=1}^{\infty} I_j(r) \left(\frac{2}{r}\right)^j (z^i a^j +\bar{a}^j z^{-j})\right)
\end{align}
where $r=\sqrt{x_1^2+x_2^2}$.
We extractaed the coefficient of the $z^0$ term, $A_0$ and it comes out to be
\begin{align}
A_0 &= I_0(\beta D/2) I_0(r) +\sum_{j=1}^{\infty} I_j(\beta D/2) I_{2j}(r) \nonumber\\  &\left(\frac{2}{r}\right)^{2 j} \left((z_0 \bar{a})^{2 j} +(z_0^{-1} a)^{2 j} \right)
\label{a01}
\end{align}
where recall that $z_0= e^{-i \alpha}$ and $\bar{a}=(x_1+ix_2)/2$. We define $\phi$ such that $\tan \phi = x_2/x_1$. Then,
\begin{align}
(z_0 \bar{a})^{2 j} +(z_0^{-1} a)^{2 j}
&= \left(\frac{r}{2}\right)^{2 j} 2 \cos 2 j (\phi-\alpha)
\end{align}
Substituting in Eq. \ref{a01}, we get
\begin{equation}
A_0 = I_0(\beta D/2) I_0(r) +2 \sum_{j=1}^{\infty} I_j(\beta D/2) I_{2 j}(r)
\cos 2j(\phi-\alpha)
\end{equation}
Since $\tilde {N} = 2 \pi e^{\beta D/2} I_0(\beta D/2)$, we get
\begin{equation}
S= I_0(r) +2 \sum_{j=1}^{\infty} \frac{I_j(\beta D/2)}{I_0(\beta D/2)} I_{2 j}(r) \cos 2j(\phi-\alpha)
\end{equation}

\section{}
\begin{figure}
     \begin{subfigure}[b]{0.2\textwidth}
         \centering
         \includegraphics[width=\textwidth]{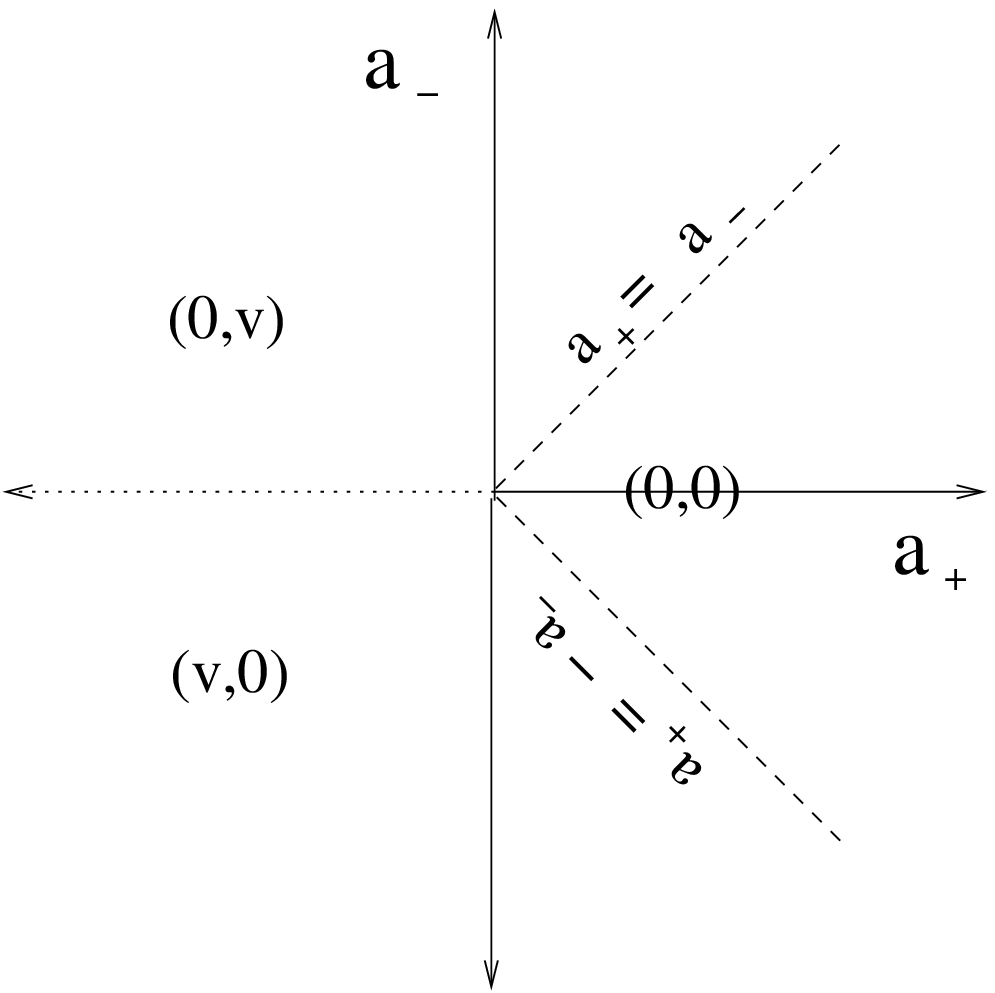}
         \caption{\footnotesize{$s_1>1$;$s_2>1$}}
         \label{bc}
     \end{subfigure}
     \hfill
     \begin{subfigure}[b]{0.26\textwidth}
         \includegraphics[width=\textwidth]{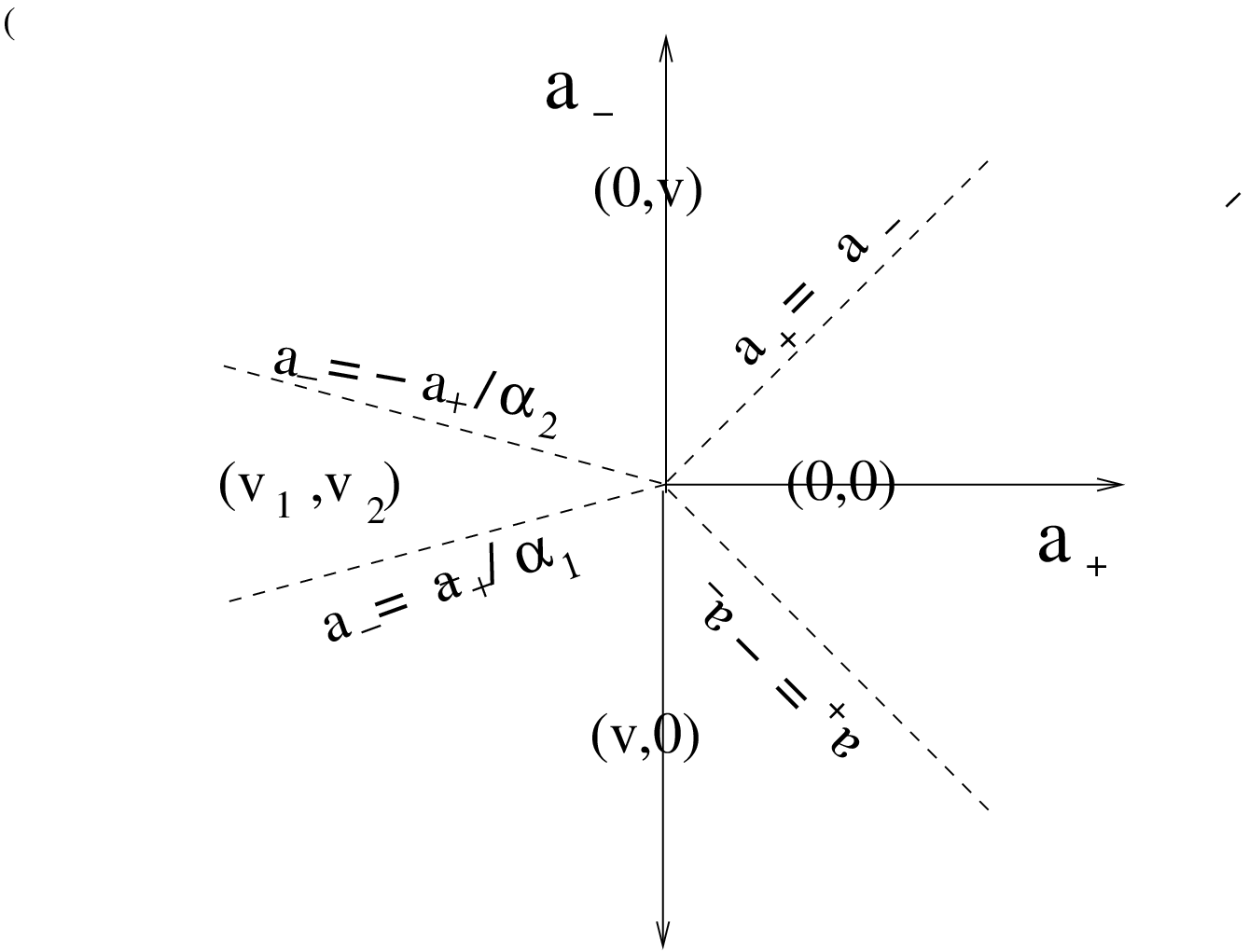}
         \caption{\scriptsize{$s_1<1$;$s_2<1$}}
         \label{te1}
     \end{subfigure}
     \hfill
     \begin{subfigure}[b]{0.22\textwidth}
         \includegraphics[width=\textwidth]{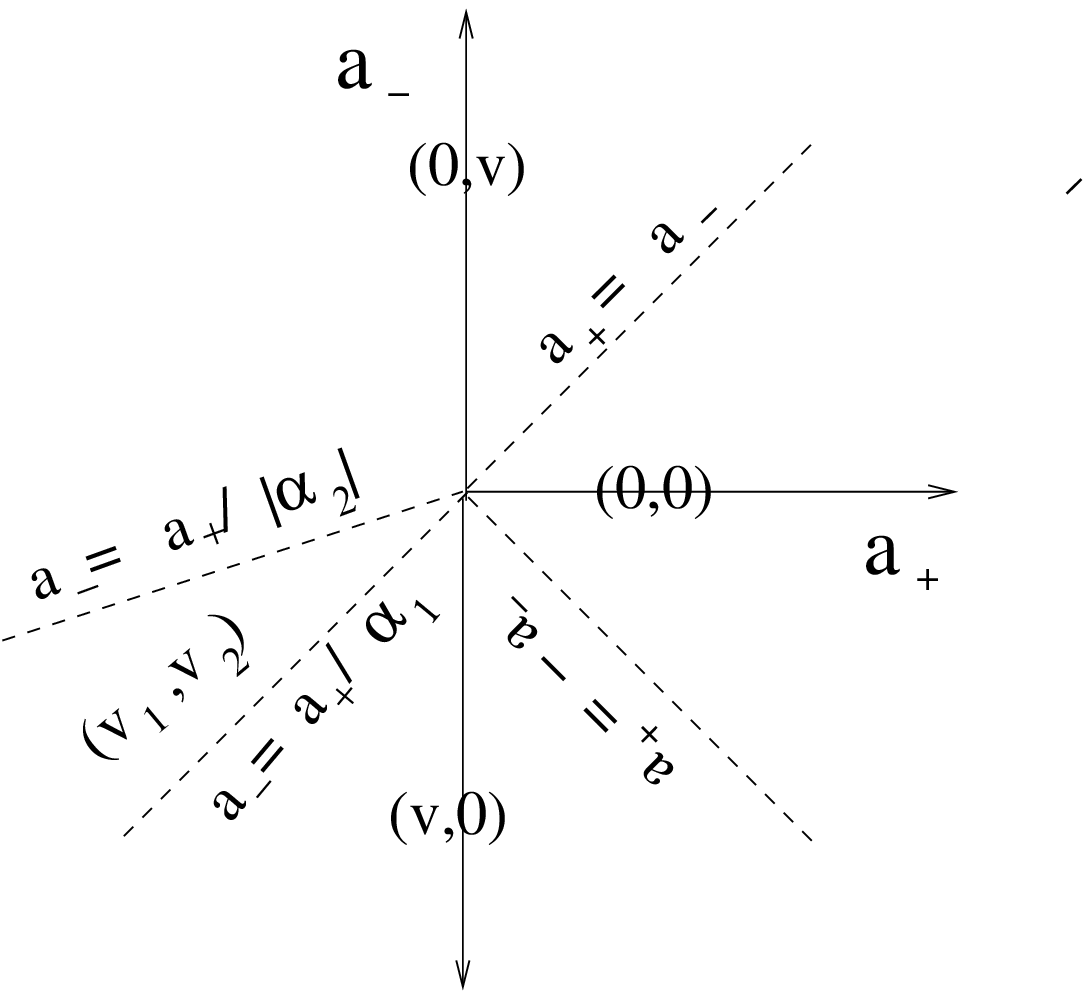}
         \caption{\scriptsize{$s_1<1$;$s_2>1$}}
         \label{te2}
     \end{subfigure}
     \begin{subfigure}[b]{0.23\textwidth}
         \includegraphics[width=\textwidth]{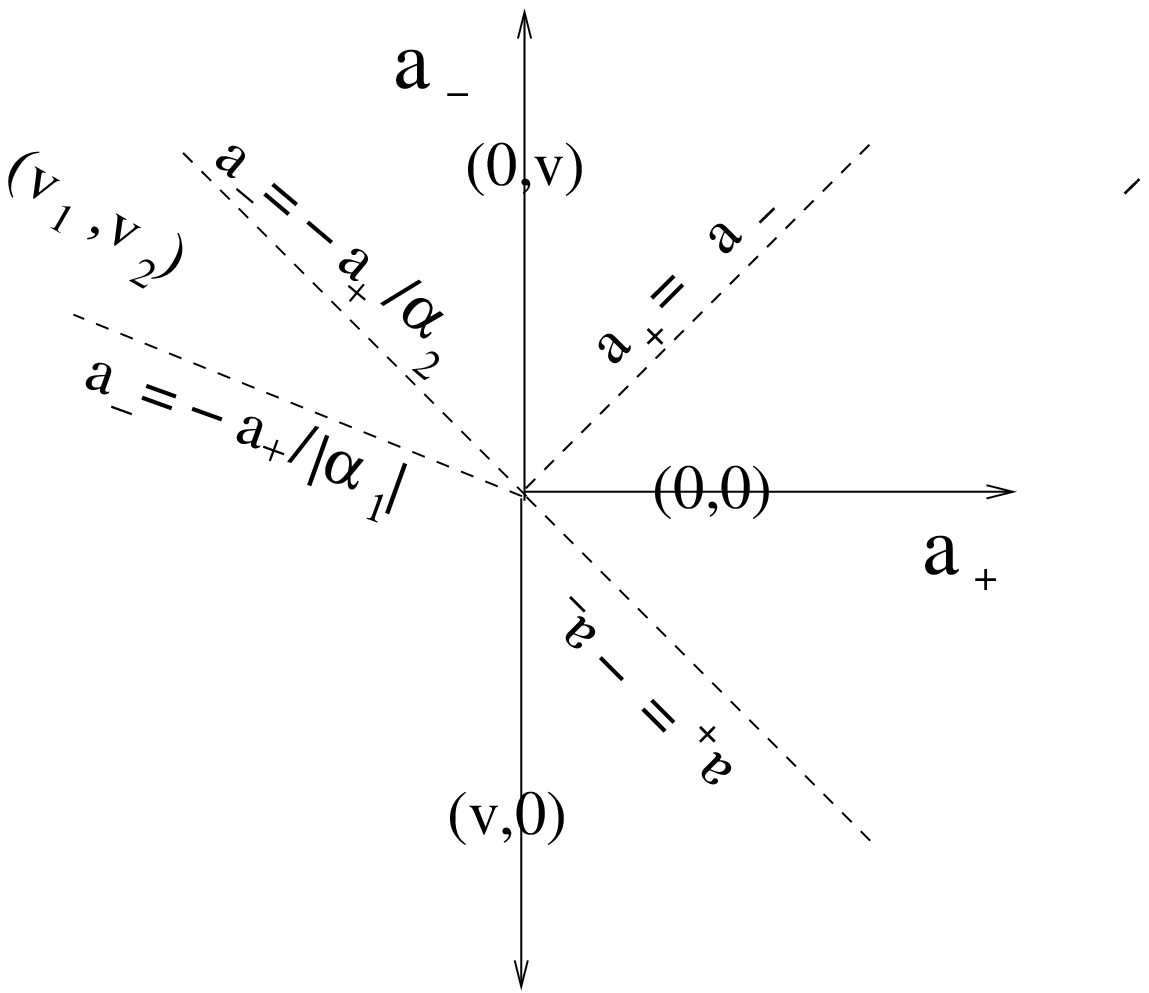}
         \caption{\scriptsize{$s_1>1$;$s_2<1$}}
         \label{te3}
     \end{subfigure}
     \hfill
             \caption{Possible phase diagrams from a two parameter Landau theory. Solid line represent the co-ordinate axes. Dashed and dotted lines represent locus of continuous and first order transitions respectively. For $u_1 u_2 < u_{12}^2$, the phase diagram consists of two lines of continuous transition and a line of first order transition meeting at a bicritical point as shown in (a). For $u_1 u_2 > u_{12}^2$ there are four phases, which meet at a tetra-critical point as shown in (b), (c) and (d). The topology though can be different depending on the ratios of $\frac{u_{12}}{u_1}$ and $\frac{u_{12}}{u_2}$.}
\label{fig:appendixb}            
\end{figure}

In this appendix we study different types of phase diagrams found with the two parameter Landau functional given in Eq. \ref{lfbd} :
\begin{align}
L(x_1,x_2) &= a_+ (x_1^2+x_2^2) + a_- (x_1^2-x_2^2)+u_1 x_1^4\nonumber\\ &+u_2 x_2^4+2 u_{12} x_1^2 x_2^2
\label{lfe}
\end{align}
where $x_1$ and $x_2$ are the two components of the order parameter and $a_+,a_-,u_1,u_2$ and $u_{12}$ are the coefficients, such that $u_1,u_2, u_{12} \geq 0$.

We observe that there are four possible states : $(0,0)$, $(0,v)$, $(v,0)$ and $(v_1,v_2)$. The fixed points $(x_1,x_2)$ of Eq. \ref{lfe} are obtained by equationg the first derivative of $L(x_1,x_2)$ w.r.t $x_1$ and $x_2$ to $0$ and satisfy the following equations
\begin{equation}
\label{fp1}
(a_++a_-) x_1+ 2 u_1 x_1^3+2 u_{12} x_1 x_2^2=0
\end{equation}
\begin{equation}
\label{fp2}
(a_+-a_-)x_2+2u_2 x_2^3+ 2 u_{12} x_1^2x_2=0
\end{equation}
The stability of states can be determined by examining the Hessian at a given fixed point. The $(i,j)^{th}$ element of the Hessian matrix is $\partial^2 L/\partial x_i \partial x_j$. The eigenvalues of the Hessian matrix for a stable state should be $\ge 0$. 

The general Hessian matrix for $L(x_1,x_2)$ is
\begin{align}
&M_H = \nonumber\\ &
\begin{pmatrix}
a_+ + a_-+6 u_1 x_1^2+2 u_{12} x_2^2 & 4 u_{12} x_1 x_2  \\
4 u_{12} x_1 x_2 & a_+-a_-+6 u_2x_2^2+2 u_{12} x_1^2  \\
\end{pmatrix}
\end{align}
There are four possible states. The region of stability of these four states is obtained by using the condition on the eigenvalues of the Hessain as follows :

{\it{Paramagnetic phase {$(x_1,x_2)=(0,0)$}}} : For this state Eq. \ref{fp1} and \ref{fp2} are trivially staisfied. The Hessian is diagonal and the conditions for both eigen values to be positive are: $(a_+ +a_-) \ge 0$ and $(a_+-a_-) \ge 0$.

{\it{{Longitudinal Ising phase $(x_1,x_2)=(v,0)$}}}: Fixed point equations are satisfied if
\begin{equation}
v^2= -\frac{a_+ +a_-}{2 u_1}
\end{equation}
The phase is stable if $(a_++a_-) \le 0$ and $a_+ (1-\frac{u_1}{u_{12}}) \ge a_- (1+\frac{u_1}{u_{12}})$.

{\it{{Transverse Ising phase $(x_1,x_2)=(0,v)$}}} : Fixed point equations are satisfied if
\begin{equation}
v^2=-\frac{(a_+-a_-)}{2 u_2}
\end{equation}
The phase is stable if : $(a_+-a_-) \le 0$ and $a
_+ (1-\frac{u_2}{u_{12}}) \le a_- (1+\frac{u_2}{u_{12}})$.

{\it{{Mixed Phase $(x_1,x_2)=(v_1,v_2)$}}} : Expression of $v_1^2$ and $v_2^2$ from the fixed point equations is 
\begin{equation}
\label{v2v1}
v_2^2 = -\frac{(a_+ +a_-)+2 u_1 v_1^2}{2 u_{12}} 
\end{equation}
\begin{equation}
\label{v1v2}
v_1^2 = -\frac{(a_+-a_-)-2 u_2 v_2^2}{2 u_{12}}
\end{equation}
The eigenvalues are:
\begin{equation}
\lambda_{\pm} = \frac{1}{2} [(u_1 v_1^2+u_2 v_2^2) \pm \sqrt{(u_1v_1^2-u_2 v_2^2)^2+4 v_1^2 v_2^2 u_{12}^2}]
\end{equation}
Both eigenvalues are greater than equal to zero when
\begin{equation}
\label{tetrac}
\frac{u_1 u_2}{u_{12}^2} \ge 1
\end{equation}
Thus if $\frac{u_1 u_2}{u_{12}^2} <1$, then there cannot be a mixed state in the system.

Besides the above condition, it also required that $v_1^2 \ge 0$ and $v_2^2 \ge 0$. Solving Eq. \ref{v2v1} and \ref{v1v2}, we get 
\begin{equation}
v_1^2 = \frac{a_+(u_{12}-u_2)-a_-(u_{12}+u_2)}{2(u_1 u_2 - u_{12}^2)}
\end{equation}
\begin{equation}
v_2^2 = \frac{a_+(u_{12}-u_1)+a_-(u_{12}+u_1)}{2(u_1 u_2 - u_{12}^2)}
\end{equation}

Since $v_1^2 \ge 0$ and $v_2^2 \ge 0$, we obtain two more additional conditions for the existence of the mixed phase : 1) $a_- (1+u_{12}/u_1) \ge a_+ (1-u_{12}/u_1)$; and 2) $a_+(u_{12}/u_2-1) \ge a_{-} (1+u_{12}/u_2)$

We define $s_1=u_{12}/u_1 $, $s_2=u_{12}/u_2$, $\alpha_1 =\frac{1+u_{12}/u_{1}}{1-u_{12}/u_1}= \frac{1+s_1}{1-s_1}$ and $\alpha_2 = \frac{1+u_{12}/u_2}{1-u_{12}/u_2}=\frac{1+s_2}{1-s_2}$. Then the  condition for existence of mixed phase is : $a_+ \le \alpha_1 a_-$ and $a_+ \le -\alpha_2 a_-$.

The Landau functional defined in Eq. \ref{lfe}, yields four different kind of phase diagrams which are described below. 

For $u_1 u_2 \le u_{12}^2$ the state $(v_1,v_2)$ is not possible. There are three states in the system and the phase diagram in $(a_+,a_-)$ plane has a bicritical point at $(0,0)$, there is a first order line along the negative $x$-axis starting at the bicritical point, separating the $(0,v)$ and $(v,0)$ phases. The $(0,v)$ and $(v,0)$ phase are separated from the $(0,0)$ phase via line of critical points along $a_+=a_-$ and $a_+=-a_-$ respectively as shown in Fig. \ref{bc}.

For $u_1 u_2>  u_{12}^2$, there are four critical lines: $a_+=a_-$, $a_+ =-a_-$, $a_+ =\alpha_1 a_-$ and $ a_+ = -\alpha_2 a_-$, which meet at $a_+=a_-=0$ in the $(a_+,a_-)$ plane. The phase $(0,0)$ exists between  the lines $a_+=a_-$ and $a_+=-a_-$. There are three different phase diagrams depending on the value of $s_1$ and $s_2$:

\begin{itemize}
\item $s_1<1$ and $s_2<1$ ( $\alpha_1>1$ and $\alpha_2>1$) : In this case $\alpha_1$ and $\alpha_2$ are both greater than one and the mixed phase occurs between $a_-=a_+/{\alpha_1}$ and $a_-= -a_+/{\alpha_2}$ as shown in Fig. \ref{te1}
\item $s_1<1$ and $s_2>1$ ($\alpha_1>1$ and $\alpha_2=- 
|\alpha_2|$, where $|\alpha_2|>1$) : In this case $\frac{|\alpha_2|}{\alpha_1} >1$ and the mixed phase exists for $a_- \ge \frac{a_+}{\alpha_1}$ and $a_- \le \frac{a_+}{|\alpha_2|}$. The phase diagram 
is as shown in Fig. \ref{te2}.
\item $s_1>1$ and $s_2<1$ ($\alpha_2>1$ and $\alpha_1=- |\alpha_1|$, where $|\alpha_1|>1$) : In this case $\frac{|\alpha_1|}{\alpha_2}>1$ and the mixed phase exists between $a_- \le \frac{-a_+}{\alpha_2} $ and $a_- \ge \frac{-a_+}{|\alpha_1|}$. The phase diagram is as shown in Fig. \ref{te3}.
\end{itemize}

We remark that the condition Eq. \ref{tetrac} for the existence of the mixed state was known earlier \cite{chaikin, watanabe}.  Besides reproducing the relation, we have shown above that the phase diagram depends also on the ratios $\frac{u_{12}}{u_1}$ and $\frac{u_{12}}{u_2}$. Here we considered only the case with $u_{12},u_1,u_2 \ge 0$; negative values of $u_{12}$ have been considered in \cite{watanabe}.


\begin{thebibliography}{99}
\bibitem{harris}R. Harris, M Plischke, J. Zuckerman, Phys. Rev. Lett. {\bf 31},160 (1973).
\bibitem{krey} U. Krey, J. of Mag. and Mag. Mat. {\bf 6}, 27 (1977).
\bibitem{alben} R. Alben, J. J. Becker and M. C. Chi, J of App. Phys. {\bf 49(3)}, 1653 (1978).
\bibitem{herzer} G. Herzer (2005) The Random Anisotropy Model. In: Idzikowski B., Švec P., Miglierini M. (eds) Properties and Applications of Nanocrystalline Alloys from Amorphous Precursors. NATO Science Series (Series II: Mathematics, Physics and Chemistry), vol 184. Springer, Dordrecht.
\bibitem{dudka} M. Dudka, R.Folk and Yu. Holovatch, J. of Magnetism and Magnetic Materials {\bf 294}, 305 (2005).
\bibitem{nano} A. Hernando, M. Vazquez, T. Kullik and C. Prados, Phys. Rev. B {\bf 51}, 3581 (1995).
\bibitem{alvarez} K. L. Alvarez, J. M. Martin, N. Burgos, M.  Ipatov, L. Dominguez and J. Gonzalez, Nanomaterials {\bf 10}, 884 (2020).
\bibitem{molecular} M. A. Girtu, C.M. Wynn, J. Zhang, J.S.
Miller, A.J. Epstein, Phys. Rev. B {\bf 61}, 492 (2000).
\bibitem{lubensky} J. H. Chen and T. C. Lubensky, Phys. Rev. B {\bf 16}, 2106 (1977).
\bibitem{bray} A. J. Bray and M. A. Moore, J Phys. C : Solid State Phys. {\bf 18}, L139 (1985).
\bibitem{callen} E. Callen, Y. J. Liu and J. R. Cullen, Phys. Rev. B {\bf 16}, 263 (1977).
\bibitem{patterson} J. D. Patterson, G. R. Gruzalski and D. J. Selimeyer, Phys. Rev. B {\bf 18}, 1377 (1978).
\bibitem{dv} B. Derrida and J. Vannimenus, J. Phys. C:Sold State Phys. {\bf 13},3261 (1980).
\bibitem{variational} D. C. Carvalho, L. M. Castro and J. A.  Plascak, Physica A {\bf 391}, 1149 (2012).
\bibitem{pelcovits} R. A. Pelcovits Phys. Rev. B {\bf 19}, 465 (1979).
\bibitem{mukamel} D. Mukamel and G. Grinstein, Phys. Rev. B,{\bf 25}, 381 (1981).
\bibitem{shapoval} D. Shapoval, M. Dudka, A. A. Fedorenko and Yu. Holovatch, Phys. Rev. B {\bf 101}, 064402 (2020).
\bibitem{referee} We thank the referee for pointing this out.
\bibitem{rg} D. Mouhanna and G. Tarjus, Phys. Rev. B {\bf 94},  214205 (2016).
\bibitem{simulation1} R. T. S. Freire and J. A. Plascak, Phys. Rev. E {\bf 91}, 032146 (2015).
\bibitem{simulation2} B. Dieny and B. Barbara, Phys. Rev. B {\bf 41}, 11549 (1990).
\bibitem{simulation3} R. Fisch, Phys. Rev. B {\bf 79}, 214429 (2009).
\bibitem{marinari} F. Liers, J. Lukic, E. Marinari, A. Pelisetto and E. Vicari, Phys. Rev. B {\bf 76},174423 (2007).
\bibitem{jayaprakash} C .Jayaprakash and S. Kirkpatrick, Phys. Rev. B {\bf 21}, 4072 (1980).
\bibitem{fischer} K. H. Fischer and A. Zippelius, J. Phys. C:Sold State Phys. {\bf 18}, L1139 (1985).
\bibitem{fisch} R. Fisch, Phys. Rev. Letts. {\bf 66},2041(1991).
\bibitem{dembo} A. Dembo and O. Zeitouni, 1998, Large Deviations Techniques and Applications (Springer-Verlag New York, Inc.). 
\bibitem{touchette} H. Touchette, Physics Reports {\bf 478}, 1 (2009).
\bibitem{lowe} M. Lowe, R. Meiners and F. Torres, J. Phys. A: Math and Theo. {\bf 46}, 125004 (2013).
\bibitem{sumedhasingh} Sumedha, and S. K. Singh, Physica A {\bf 442}, 276 (2016).
\bibitem{sumedhajana} Sumedha, and N. K. Jana, J. Phys A: Math and Theo.  {\bf 50}, 015003 (2017).
\bibitem{kistler} L. P. Arguin and N. Kistler, J. Stat. Phys. {\bf 157}, 1 (2014).
\bibitem{kirkpatrick} K. Kikrpatrick and T. Nawaz, J. Stat. Phys.  {\bf 165}, 1114 (2016).
\bibitem{sumedhabarma} Sumedha, and M. Barma, J. Phys A: Math and Theo. (2022), https://doi.org/10.1088/1751-8121/ac4b8b, arXiv:2104.06664.
\bibitem{kosterlitz} J.  M. Kosterlitz, D. R. Nelson and M. E. Fisher, Phys. Rev. B {\bf 13}, 412 (1976).
\bibitem{aharony} A. Aharony and S. Fishman, Phys. Rev. Letts. {\bf 37}, 1587 (1976).
\bibitem{zhang} S. C. Zhang, Science {\bf 275}, 1089 (1997).
\bibitem{murakami} S. Murakami and N. Nagaosa, Journal of the Physical Society of Japan, {\bf 69}, 2395-2398 (2000). 
\bibitem{sannino} F. Sannino and K. Touminen, Phys. Rev. D, {\bf 70}, 034019 (2004).
\bibitem{hollander} Frank den Hollander, Large Deviations, Fields Institute Monographs, AMS (2000) Theorem III.17.
\bibitem{abramowitz} M. Abramowitz, 1974, Handbook of Mathematical Functions, with Formulas, Graphs, and Mathematical Tables (Dover Publications, New York).
\bibitem{chaikin} P. M. Chaikin and T. C. Lubensky, 2000, Principles of Condensed Matter Physics (Cambridge University Press, Cambridge).
\bibitem{watanabe} {S. Watanabe and T. Usui, Prog. Theor. Phys. {\bf{73}}, 1305 (1985).}


\end{thebibliography}
\end{document}